\documentstyle[10pt,epsf,dp_delphititle,multirow,epsfig,float]{dp_delphi}
\makeindex
\def\DpPaperGroup{PH-EP}
\def\DpPaperRef{2005-026}
\def\DpDate{20 May 2005}
\def\DpAuthors{DELPHI Collaboration}
\def\DpSubmit{(Eur. Phys. J. C45 (2006) 273-289)}
\def\DpTitle{{
Single Intermediate Vector Boson \\ 
production in ${\boldmath e^+e^-}$ collisions \\
at $\boldmath{\sqrt{s} = 183 - 209}$ GeV
}}
\def\DpComment{}
\def\DpEMail{}

\def\lu{{\cal {L}}_{int}}
\def\pbi{pb$^{-1}$}
\def\epem{e^+e^-}
\def\qq{q\bar{q}}
\def\mm{\mu^+ \mu^-}

\def\eell{e^+e^-\to l^+l^-}

\def\eeww{e^+e^-\to W^+W^-}
\def\eezz{e^+e^-\to ZZ}

\def\eewe{e^+e^-\to e^- \bar{\nu}_e W^+}

\def\eegze{e^+e^-\to e^+e^-\gamma^*/Z}
\def\gewv{\gamma e^+\to \bar{\nu}_e W^+}
\def\geze{\gamma e^+\to e^+ Z}

\def\wev{e\nu_eW}
\def\evqq{e\nu_e{q}\bar{q}'}

\def\evlv{e\nu_el\nu_l}
\def\evmv{e\nu_e\mu\nu_\mu}
\def\evtv{e\nu_e\tau\nu_\tau}
\def\evev{e\nu_ee\nu_e}
\def\gs{\gamma^{*}}
\def\deg{^{\circ}}
\def\qbar{\bar{q}}
\def\fbar{\bar{f}}
\begin{document}
\makeatletter
\newcount\@tempcntc
\def\@citex[#1]#2{\if@filesw\immediate\write\@auxout{\string\citation{#2}}\fi
\@tempcnta\z@\@tempcntb\m@ne\def\@citea{}\@cite{\@for\@citeb:=#2\do
{\@ifundefined
{b@\@citeb}{\@citeo\@tempcntb\m@ne\@citea\def\@citea{,}{\bf ?}\@warning
{Citation `\@citeb' on page \thepage \space undefined}}%
{\setbox\z@\hbox{\global\@tempcntc0\csname b@\@citeb\endcsname\relax}%
\ifnum\@tempcntc=\z@ \@citeo\@tempcntb\m@ne
\@citea\def\@citea{,}\hbox{\csname b@\@citeb\endcsname}%
\else
\advance\@tempcntb\@ne
\ifnum\@tempcntb=\@tempcntc
\else\advance\@tempcntb\m@ne\@citeo
\@tempcnta\@tempcntc\@tempcntb\@tempcntc\fi\fi}}\@citeo}{#1}}
\def\@citeo{\ifnum\@tempcnta>\@tempcntb\else\@citea\def\@citea{,}%
\ifnum\@tempcnta=\@tempcntb\the\@tempcnta\else
{\advance\@tempcnta\@ne\ifnum\@tempcnta=\@tempcntb \else \def\@citea{--}\fi
\advance\@tempcnta\m@ne\the\@tempcnta\@citea\the\@tempcntb}\fi\fi}
\makeatother
\begin{titlepage}
\pagenumbering{roman}
\CERNpreprint{\DpPaperGroup}{\DpPaperRef} 
\date{{\small\DpDate}} 
\title{\DpTitle} 
\address{\DpAuthors} 
\begin{shortabs} 
\noindent
%
\noindent
%
\noindent
The production of single charged and neutral intermediate vector
bosons in $e^+ e^-$ collisions has been studied in the data collected
by the DELPHI experiment at LEP at centre-of-mass energies between 183
and 209 GeV, corresponding to an integrated luminosity of about
640~pb$^{-1}$.    
The measured cross-sections for the reactions, determined in limited
kinematic regions, are in agreement with the Standard Model predictions.

\end{shortabs}
\vfill
\begin{center}
\DpSubmit \ \\ 
\DpComment \ \\
\DpEMail \ \\
\end{center}
\vfill
\clearpage
\headsep 10.0pt
\addtolength{\textheight}{10mm}
\addtolength{\footskip}{-5mm}
\begingroup
%
\newcommand{\DpName}[2]{\hbox{#1$^{\ref{#2}}$},\hfill}
\newcommand{\DpNameTwo}[3]{\hbox{#1$^{\ref{#2},\ref{#3}}$},\hfill}
\newcommand{\DpNameThree}[4]{\hbox{#1$^{\ref{#2},\ref{#3},\ref{#4}}$},\hfill}
\newskip\Bigfill \Bigfill = 0pt plus 1000fill
\newcommand{\DpNameLast}[2]{\hbox{#1$^{\ref{#2}}$}\hspace{\Bigfill}}
%
\footnotesize
\noindent
\DpName{J.Abdallah}{LPNHE}
\DpName{P.Abreu}{LIP}
\DpName{W.Adam}{VIENNA}
\DpName{P.Adzic}{DEMOKRITOS}
\DpName{T.Albrecht}{KARLSRUHE}
\DpName{T.Alderweireld}{AIM}
\DpName{R.Alemany-Fernandez}{CERN}
\DpName{T.Allmendinger}{KARLSRUHE}
\DpName{P.P.Allport}{LIVERPOOL}
\DpName{U.Amaldi}{MILANO2}
% ...
\normalsize
\endgroup
\titlefoot{Department of Physics and Astronomy, Iowa State
     University, Ames IA 50011-3160, USA
    \label{AMES}}
\titlefoot{Physics Department, Universiteit Antwerpen,
     Universiteitsplein 1, B-2610 Antwerpen, Belgium \\
     \indent~~and IIHE, ULB-VUB,
     Pleinlaan 2, B-1050 Brussels, Belgium \\
     \indent~~and Facult\'e des Sciences,
     Univ. de l'Etat Mons, Av. Maistriau 19, B-7000 Mons, Belgium
    \label{AIM}}
\titlefoot{Physics Laboratory, University of Athens, Solonos Str.
     104, GR-10680 Athens, Greece
    \label{ATHENS}}
\titlefoot{Department of Physics, University of Bergen,
     All\'egaten 55, NO-5007 Bergen, Norway
    \label{BERGEN}}
\titlefoot{Dipartimento di Fisica, Universit\`a di Bologna and INFN,
     Via Irnerio 46, IT-40126 Bologna, Italy
    \label{BOLOGNA}}
\titlefoot{Centro Brasileiro de Pesquisas F\'{\i}sicas, rua Xavier Sigaud 150,
     BR-22290 Rio de Janeiro, Brazil \\
     \indent~~and Depto. de F\'{\i}sica, Pont. Univ. Cat\'olica,
     C.P. 38071 BR-22453 Rio de Janeiro, Brazil \\
     \indent~~and Inst. de F\'{\i}sica, Univ. Estadual do Rio de Janeiro,
     rua S\~{a}o Francisco Xavier 524, Rio de Janeiro, Brazil
    \label{BRASIL}}
\titlefoot{Coll\`ege de France, Lab. de Physique Corpusculaire, IN2P3-CNRS,
     FR-75231 Paris Cedex 05, France
    \label{CDF}}
\titlefoot{CERN, CH-1211 Geneva 23, Switzerland
    \label{CERN}}
\titlefoot{Institut de Recherches Subatomiques, IN2P3 - CNRS/ULP - BP20,
     FR-67037 Strasbourg Cedex, France
    \label{CRN}}
\titlefoot{Now at DESY-Zeuthen, Platanenallee 6, D-15735 Zeuthen, Germany
    \label{DESY}}
\titlefoot{Institute of Nuclear Physics, N.C.S.R. Demokritos,
     P.O. Box 60228, GR-15310 Athens, Greece
    \label{DEMOKRITOS}}
\titlefoot{FZU, Inst. of Phys. of the C.A.S. High Energy Physics Division,
     Na Slovance 2, CZ-180 40, Praha 8, Czech Republic
    \label{FZU}}
\titlefoot{Dipartimento di Fisica, Universit\`a di Genova and INFN,
     Via Dodecaneso 33, IT-16146 Genova, Italy
    \label{GENOVA}}
\titlefoot{Institut des Sciences Nucl\'eaires, IN2P3-CNRS, Universit\'e
     de Grenoble 1, FR-38026 Grenoble Cedex, France
    \label{GRENOBLE}}
\titlefoot{Helsinki Institute of Physics and Department of Physical Sciences,
     P.O. Box 64, FIN-00014 University of Helsinki, 
     \indent~~Finland
    \label{HELSINKI}}
\titlefoot{Joint Institute for Nuclear Research, Dubna, Head Post
     Office, P.O. Box 79, RU-101 000 Moscow, Russian Federation
    \label{JINR}}
\titlefoot{Institut f\"ur Experimentelle Kernphysik,
     Universit\"at Karlsruhe, Postfach 6980, DE-76128 Karlsruhe,
     Germany
    \label{KARLSRUHE}}
\titlefoot{Institute of Nuclear Physics PAN,Ul. Radzikowskiego 152,
     PL-31142 Krakow, Poland
    \label{KRAKOW1}}
\titlefoot{Faculty of Physics and Nuclear Techniques, University of Mining
     and Metallurgy, PL-30055 Krakow, Poland
    \label{KRAKOW2}}
\titlefoot{Universit\'e de Paris-Sud, Lab. de l'Acc\'el\'erateur
     Lin\'eaire, IN2P3-CNRS, B\^{a}t. 200, FR-91405 Orsay Cedex, France
    \label{LAL}}
\titlefoot{School of Physics and Chemistry, University of Lancaster,
     Lancaster LA1 4YB, UK
    \label{LANCASTER}}
\titlefoot{LIP, IST, FCUL - Av. Elias Garcia, 14-$1^{o}$,
     PT-1000 Lisboa Codex, Portugal
    \label{LIP}}
\titlefoot{Department of Physics, University of Liverpool, P.O.
     Box 147, Liverpool L69 3BX, UK
    \label{LIVERPOOL}}
\titlefoot{Dept. of Physics and Astronomy, Kelvin Building,
     University of Glasgow, Glasgow G12 8QQ
    \label{GLASGOW}}
\titlefoot{LPNHE, IN2P3-CNRS, Univ.~Paris VI et VII, Tour 33 (RdC),
     4 place Jussieu, FR-75252 Paris Cedex 05, France
    \label{LPNHE}}
\titlefoot{Department of Physics, University of Lund,
     S\"olvegatan 14, SE-223 63 Lund, Sweden
    \label{LUND}}
\titlefoot{Universit\'e Claude Bernard de Lyon, IPNL, IN2P3-CNRS,
     FR-69622 Villeurbanne Cedex, France
    \label{LYON}}
\titlefoot{Dipartimento di Fisica, Universit\`a di Milano and INFN-MILANO,
     Via Celoria 16, IT-20133 Milan, Italy
    \label{MILANO}}
\titlefoot{Dipartimento di Fisica, Univ. di Milano-Bicocca and
     INFN-MILANO, Piazza della Scienza 2, IT-20126 Milan, Italy
    \label{MILANO2}}
\titlefoot{IPNP of MFF, Charles Univ., Areal MFF,
     V Holesovickach 2, CZ-180 00, Praha 8, Czech Republic
    \label{NC}}
\titlefoot{NIKHEF, Postbus 41882, NL-1009 DB
     Amsterdam, The Netherlands
    \label{NIKHEF}}
\titlefoot{National Technical University, Physics Department,
     Zografou Campus, GR-15773 Athens, Greece
    \label{NTU-ATHENS}}
\titlefoot{Physics Department, University of Oslo, Blindern,
     NO-0316 Oslo, Norway
    \label{OSLO}}
\titlefoot{Dpto. Fisica, Univ. Oviedo, Avda. Calvo Sotelo
     s/n, ES-33007 Oviedo, Spain
    \label{OVIEDO}}
\titlefoot{Department of Physics, University of Oxford,
     Keble Road, Oxford OX1 3RH, UK
    \label{OXFORD}}
\titlefoot{Dipartimento di Fisica, Universit\`a di Padova and
     INFN, Via Marzolo 8, IT-35131 Padua, Italy
    \label{PADOVA}}
\titlefoot{Rutherford Appleton Laboratory, Chilton, Didcot
     OX11 OQX, UK
    \label{RAL}}
\titlefoot{Dipartimento di Fisica, Universit\`a di Roma II and
     INFN, Tor Vergata, IT-00173 Rome, Italy
    \label{ROMA2}}
\titlefoot{Dipartimento di Fisica, Universit\`a di Roma III and
     INFN, Via della Vasca Navale 84, IT-00146 Rome, Italy
    \label{ROMA3}}
\titlefoot{DAPNIA/Service de Physique des Particules,
     CEA-Saclay, FR-91191 Gif-sur-Yvette Cedex, France
    \label{SACLAY}}
\titlefoot{Instituto de Fisica de Cantabria (CSIC-UC), Avda.
     los Castros s/n, ES-39006 Santander, Spain
    \label{SANTANDER}}
\titlefoot{Inst. for High Energy Physics, Serpukov
     P.O. Box 35, Protvino, (Moscow Region), Russian Federation
    \label{SERPUKHOV}}
\titlefoot{J. Stefan Institute, Jamova 39, SI-1000 Ljubljana, Slovenia
     and Laboratory for Astroparticle Physics,\\
     \indent~~Nova Gorica Polytechnic, Kostanjeviska 16a, SI-5000 Nova Gorica, Slovenia, \\
     \indent~~and Department of Physics, University of Ljubljana,
     SI-1000 Ljubljana, Slovenia
    \label{SLOVENIJA}}
\titlefoot{Fysikum, Stockholm University,
     Box 6730, SE-113 85 Stockholm, Sweden
    \label{STOCKHOLM}}
\titlefoot{Dipartimento di Fisica Sperimentale, Universit\`a di
     Torino and INFN, Via P. Giuria 1, IT-10125 Turin, Italy
    \label{TORINO}}
\titlefoot{INFN,Sezione di Torino and Dipartimento di Fisica Teorica,
     Universit\`a di Torino, Via Giuria 1,
     IT-10125 Turin, Italy
    \label{TORINOTH}}
\titlefoot{Dipartimento di Fisica, Universit\`a di Trieste and
     INFN, Via A. Valerio 2, IT-34127 Trieste, Italy \\
     \indent~~and Istituto di Fisica, Universit\`a di Udine,
     IT-33100 Udine, Italy
    \label{TU}}
\titlefoot{Univ. Federal do Rio de Janeiro, C.P. 68528
     Cidade Univ., Ilha do Fund\~ao
     BR-21945-970 Rio de Janeiro, Brazil
    \label{UFRJ}}
\titlefoot{Department of Radiation Sciences, University of
     Uppsala, P.O. Box 535, SE-751 21 Uppsala, Sweden
    \label{UPPSALA}}
\titlefoot{IFIC, Valencia-CSIC, and D.F.A.M.N., U. de Valencia,
     Avda. Dr. Moliner 50, ES-46100 Burjassot (Valencia), Spain
    \label{VALENCIA}}
\titlefoot{Institut f\"ur Hochenergiephysik, \"Osterr. Akad.
     d. Wissensch., Nikolsdorfergasse 18, AT-1050 Vienna, Austria
    \label{VIENNA}}
\titlefoot{Inst. Nuclear Studies and University of Warsaw, Ul.
     Hoza 69, PL-00681 Warsaw, Poland
    \label{WARSZAWA}}
\titlefoot{Fachbereich Physik, University of Wuppertal, Postfach
     100 127, DE-42097 Wuppertal, Germany
    \label{WUPPERTAL}}
\addtolength{\textheight}{-10mm}
\addtolength{\footskip}{5mm}
\clearpage
\headsep 30.0pt
\end{titlepage}
%
\pagenumbering{arabic} 
\setcounter{footnote}{0} %
\large
%
\section{Introduction}

The production of four-fermion final states becomes
increasingly important in $\epem$ 
collisions at centre-of-mass energies above the $Z$ pole.
The full set of Feynman diagrams must be considered, 
but particular topologies receive their dominant contribution from a
subset of them (cf. Figure~\ref{fig:Feynman}, taken from
reference~\cite{YRlep2}, for the standard definition of the different
graphs). 
As the centre-of-mass energy increases, the dominant processes leading
to production of vector bosons are represented by the bremsstrahlung
and fusion diagrams with the production of a single vector boson\footnote{
Charge conjugate states are implied throughout the text.} ($\eewe$,
$\eegze$), and the conversion and non-Abelian annihilation diagrams
leading to doubly resonant production ($\eeww$, $\eezz$).  
Single resonant production is dominated by the bremsstrahlung process,
which proceeds through the scattering of a quasi-real 
photon ($q^2_\gamma \sim 0$) radiated from an incoming $e^-$ on an
$e^+$ of the other beam, i.e.: $\gewv$, $\geze$~\cite{singleBoson}.
The resulting topology is characterized by the $e^-$ which radiates the
quasi-real photon being predominantly lost along the beam line.  

The integrated luminosity delivered by the LEP collider in the runs at
centre-of-mass energies $\sqrt{s}=183-209$ GeV (LEP2) allowed, for
the first time, measurements of the cross-section of single
boson production and not just the observation of individual events. 
The evaluation of the Standard Model cross-sections for this process requires
the computation of the full set of Feynman diagrams and, to deal with
the collinear singularity corresponding to the electron lost along the beam
line, the use of fully massive matrix elements. 
Moreover the different scales of the
couplings in the process and the scale for the QED initial
state radiation (ISR) should be properly accounted for to
provide a reliable prediction. 
Therefore this process was taken as a benchmark when comparing the
different calculations used to describe four-fermion physics at
LEP2~\cite{4fYR}.   
In addition, because of the large missing energy in the final state,
single boson production is a background when probing
for new physics as in the search for the Higgs boson in the
$H\nu\bar{\nu}$ channel or for physics beyond the
Standard Model~\cite{beySM}.  
Therefore the measurement of its cross-section
is an important check that the background 
in these searches is correctly modelled.
By itself, single-$W$ production provides access to the measurement
of the trilinear gauge couplings at the $WW\gamma$ vertex: 
this measurement, in combination with other physics channels, has been
made by the {\mbox DELPHI} Collaboration and is reported
elsewhere~\cite{paptgc}. 
Finally, single boson production is interesting as it will be the
dominant source of weak boson production at a forthcoming Linear
Collider. 

Single boson production is investigated in this paper in five different
final states:  $e^- \bar{\nu}_e q \bar{q}'$, $e^-\bar{\nu}_e \mu^+\nu_\mu$
and $e^- \bar{\nu}_e  e^+ \nu_e$ for single-$W$
production, $e^-e^+q\bar{q}$ and $e^-e^+\mu^-\mu^+$ for single-$Z$ production. 
Cross-sections are measured using the data collected by the {\mbox DELPHI}
experiment at centre-of-mass energies ranging from 183 to
209~GeV with a corresponding integrated luminosity of about 640~\pbi. 
Compared to the results reported by the DELPHI Collaboration
in~\cite{plb01}, those  reported here are based on a larger sample, a better data
processing and an improved description of the simulated events (see below). 
Therefore they update and supersede those already reported in~\cite{plb01}. 
Results on single boson processes have been published by the other LEP
experiments in~\cite{Aleph,L3} for single-$W$ and
in~\cite{Aleph,L3,Opal} for single-$Z$ production. 

The criteria for the selection of the events are mainly based on the 
information from the tracking system, the calorimeters and the muon chambers
of the {\mbox DELPHI} detector. 
A detailed description of the {\mbox DELPHI} apparatus and its
performance can be found in~\cite{DELPHI}. The detector
has remained essentially unchanged in the LEP2 phase, except for
upgrades of the Vertex Detector~\cite{VD} and the addition of a set of
scintillation counters to veto photons in the blind regions of the
electromagnetic calorimetry at polar angles\footnote{In the
  reference frame used in DELPHI the $z$ axis
  was oriented along the incoming $e^-$ beam, $\theta$ indicated the polar
  angle and $\phi$ the azimuthal angle.} 
$\theta \simeq 40^\circ$,
$\theta \simeq 90^\circ$ and $\theta \simeq 140^\circ$.
The main tracking device was the Time Projection Chamber (TPC).
One of the sectors (1/12 of the azimuthal acceptance) of the TPC, hereafter
indicated as S6, was not fully operational during the last period of
data taking at $\sqrt{s} = 207$ GeV (about 50 pb$^{-1}$).   
The data with the TPC sector down were analysed separately,
with the performance of the analysis being evaluated on dedicated
simulation samples, where this effect was explicitly taken into account.

This paper is organized as follows. Section~\ref{sec:def} introduces the
definition of the single boson processes in terms of the four-fermion
final states.
In Sections~\ref{sec:singleW} and~\ref{sec:singleZ} the selection of
the events and the extraction of the single-$W$ and single-$Z$ cross-sections are
presented. 
In Section~\ref{sec:conclusions} the comparison of measured cross-sections 
with the predictions of the Standard Model is performed for all the
single boson final states.

\section{Definition of the signal and simulation}\label{sec:def}

\begin{figure}[htb]
\begin{center}
\epsfig{file=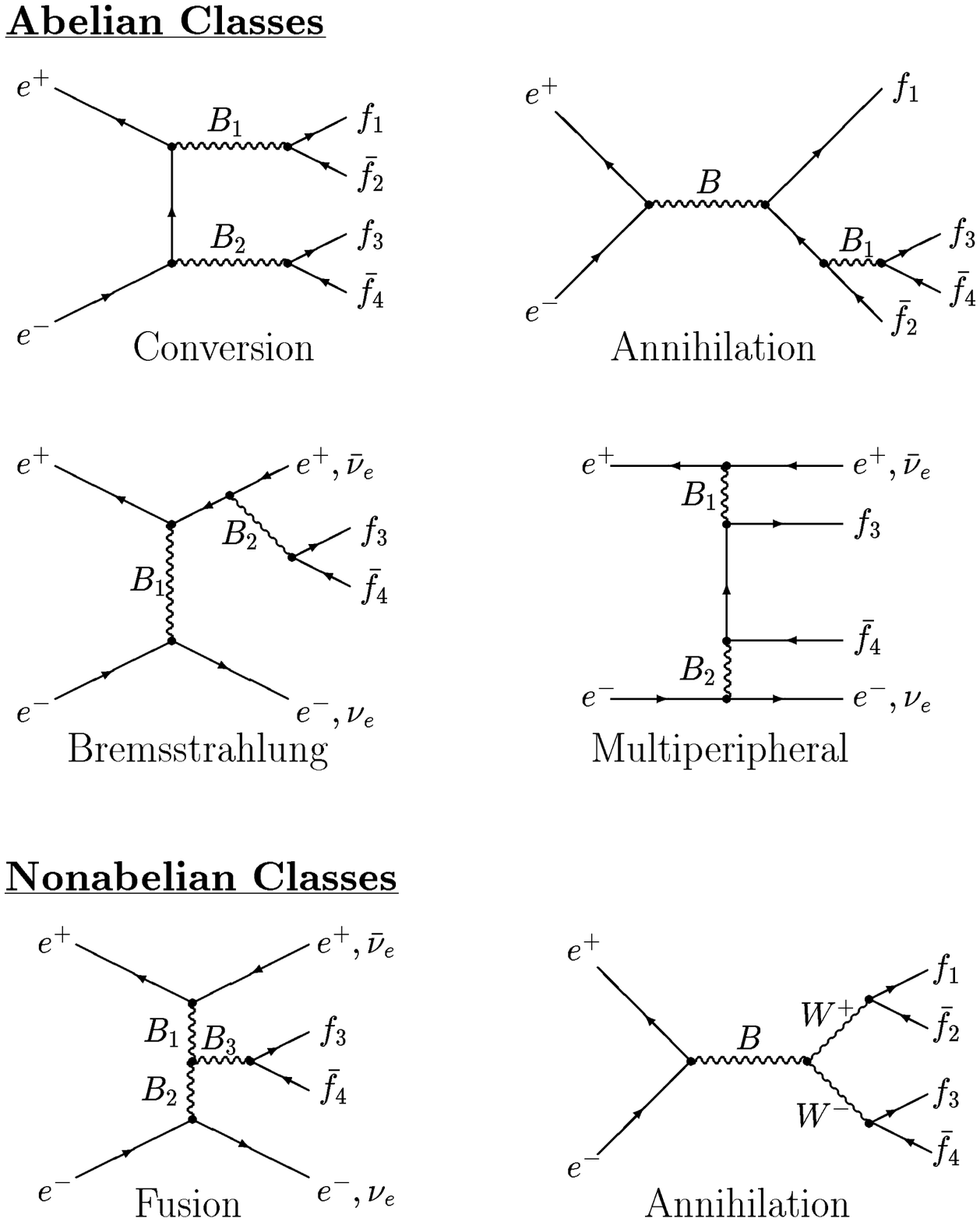,width=10cm}
\end{center}
\caption{Four-fermion production classes of diagrams in $e^+e^-$
annihilation following the convention of~\cite{YRlep2}: $B=Z,\gamma$ and
$B_1,B_2,B_3=Z,\gamma,W^{\pm}$. Diagrams involving Higgs boson
exchange are not shown.}  
\label{fig:Feynman}
\end{figure}

Single boson production is investigated in this paper through
four-fermion final states, $e^- \bar{\nu}_e f\bar{f}'$ and $e^+ e^- f\bar{f}$. 
These final states receive contributions not only from single resonant
diagrams but also from doubly resonant production, conversion
diagrams and multiperipheral processes (Figure~\ref{fig:Feynman},
according to the convention of~\cite{YRlep2}).
To enhance the single boson production contribution, and to enable
consistent comparisons and combinations between the experiments at
LEP, the cross-sections have been defined in the limited kinematic
regions described below. 

\paragraph{\mbox{\boldmath $\wev$} {\bf channel:} }
The four-fermion final states $e^- \bar{\nu}_e q \bar{q}'$ and $e^-
\bar{\nu}_e l^+ \nu_l$   ($l=\mu,\tau$) can be produced both via
single-$W$ production, referred to
as $\wev$ in the following, or via $W$-pair production.
A distinctive feature of $\wev$ is the fact that the distribution of the
electron direction is strongly peaked at small polar angles
($\theta_{e}$) with respect to the incoming electron beam direction.     
Based on this consideration, the $\wev$ signal was defined by the
complete $t$-channel subset\footnote{The $t$-channel subset consists
  of  the bremsstrahlung, fusion and multiperipheral diagrams of
Figure~\ref{fig:Feynman}.} of the Feynman diagrams contributing to the $e^-
\bar{\nu}_e q \bar{q}'$ and $e^-\bar{\nu}_e l^+ \nu_l$ final states
with additional kinematic selections to exclude the regions of the phase
space dominated by multiperipheral diagrams, where the cross-section
calculation is affected by large uncertainties. The signal region was
therefore defined as follows:
\begin{eqnarray} 
m_{q\bar{q}'} > 45 \ \mbox{GeV}/c^2 \ \ &
\mbox{for} & \ \ e^{-}\bar{\nu}_{e} q \bar{q}', \\
E_{l^+} > 20 \ \mbox{GeV} \ \  &
\mbox{for} & \ \ e^{-}\bar{\nu}_{e} l^{+}\nu_{l}
 \ \ (l^{+}= \mu^{+}, \tau^{+}), \nonumber 
\label{eq:cut2}
\end{eqnarray}
where $m_{q\bar{q}'}$ is the $q\bar{q}'$ invariant mass and $E_{l^+}$
the lepton energy.
Single-$W$ production accounts for more than $80 \%$ of all 
$e^{-}\bar{\nu}_{e} q \bar{q}'$ and $e^{-}\bar{\nu}_{e} l^{+}\nu_{l}$ 
events in the kinematic region defined above.


\paragraph{\mbox{\boldmath $e\nu_e e\nu_e$} {\bf channel:}}
In the kinematic region with one electron lost in the forward
direction, this final state receives, 
besides single-$W$ production, a large contribution from
single-$Z$ production (with $Z \to \nu_e\bar{\nu}_e$) and from the 
interference between single-$W$ and single-$Z$ processes. 
The contribution of this channel to the $\wev$ signal was also defined
by the complete $t$-channel subset of the Feynman diagrams, in this
case with the following kinematic selections: 
\begin{equation}
|\cos\theta_{e^+}| < 0.95,~~E_{e^+} > 20 ~{\mathrm GeV~and}~~|\cos\theta_{e^-}| > 0.95.
\label{eq:cut3}
\end{equation}
\noindent The $e^- \bar{\nu}_e  e^+ \nu_e$ channel was not used in the
determination of the single-$Z$ production cross-section.


\paragraph{\mbox{\boldmath $Zee$} {\bf channel:}}
The neutral bosons are produced in the process $e \gamma \to  e \gs /Z$, 
where a quasi-real photon is radiated from one of the beam electrons
and scattered off the other beam. 
The signature of such events is an electron in the detector, typically
of low energy, recoiling against the $\gs/Z$ system, with the other electron
usually lost in the beam-pipe. 
The $Zee$ cross-section, defined for the combination of the results from all the LEP
experiments, refers to the entire set of 48 graphs contributing at tree level to the
$e^+e^-f\fbar$ $(f=q,\mu)$ final state with the following restrictions
in the phase space to enhance the single boson contribution\footnote{Diagrams involving Higgs
  boson exchange are neglected. Multiperipheral processes with at least one photon resolved are excluded from 
the signal definition even if they are within the accepted kinematic region.
}:
\begin{eqnarray}
 &m_{f\fbar} > 60~\mathrm{GeV}/{\it c}^2~~\mathrm{and}& \\ \nonumber
 &\theta_{e^+}>168\deg,~~60\deg<\theta_{e^-}<168\deg~~{\mathrm and}~E_{e^-}
 > 3~\mathrm{GeV} & \mathrm{for~a~visible~electron, or} \\
 &
 \theta_{e^-}<12\deg,~~12\deg<\theta_{e^+}<120\deg~~{\mathrm and}~E_{e^+}
 > 3~\mathrm{GeV} & \mathrm{for~a~visible~positron,}   \nonumber
\end{eqnarray}
\noindent the $12\deg$ ($168\deg$) being motivated by the lower angle
of the acceptance for the electron identification of the LEP experiments.

At $\sqrt{s} = 200$ GeV, within these kinematic limits, the bremsstrahlung contribution 
accounts for about 97\% of $e^+e^-q\qbar$ and 67\% of $e^+e^-\mu^+\mu^-$ final states.
The cut on the invariant mass was set to 60 GeV/$c^2$ because it both
guarantees an efficient rejection of the multiperipheral contribution 
and it provides a natural separation between the $\gs ee$ and
$Zee$ regions, as it corresponds to the minimum of the differential
$m_{f\fbar}$ distribution. This paper presents, for the $e^+ e^-q\bar{q}$
final state, besides the above defined $Zee$ cross-section, a
measurement of the cross-section with the same acceptance cuts for
$e^+$ and $e^-$ but in the invariant mass range $15 < m_{q\qbar} < 60$
GeV/$c^2$ (hereafter referred to as $\gs ee$).

\paragraph{} The $\wev$ and $Zee$ signal samples, as predicted by the
Standard Model, were simulated with
the \mbox{WPHACT}~\cite{wphact} event generator.  
For background processes, different
generators were used: \mbox{KK2f}~\cite{kk2f} for
$q\qbar(\gamma)$, $e^+e^-\to\mu^+\mu^-(\gamma)$ and $\tau^+\tau^-(\gamma)$,
\mbox{TEEGG}~\cite{teeg} and  BHWIDE~\cite{bhwide} for
$e^+e^-\to{e^+}e^-\gamma$, \mbox{PYTHIA 6.143}~\cite{pythia} 
and BDK~\cite{bdk} for two-photon collisions. 
Fragmentation and hadronization for the \mbox{KK2f} and 
\mbox{WPHACT} samples were performed using \mbox{PYTHIA 6.156}. 
A detailed description of the simulation of four-fermion events at
LEP2 as done in DELPHI is given in~\cite{4fmc}.
The generated signal and background events were passed through the
detailed simulation of the {\mbox DELPHI} detector~\cite{DELPHI} and
then processed with the same reconstruction and analysis programs as
the data.

\section{Single-\mbox{\boldmath $W$} analysis}\label{sec:singleW}

Both the hadronic and the leptonic final states were considered in
the single-$W$ analysis. They are characterized by the presence of two
hadronic jets acoplanar with the beam or by a single lepton with large
transverse momentum, respectively. The final state electron is 
lost in the beam pipe.
 
\subsection{Selection of hadronic events}

The experimental signature of $e^-\bar{\nu}_e q\bar{q}'$ events consists of a pair
of acoplanar jets. The
undetected neutrino results in a large missing momentum at large angle
to the beam direction.

Other physics processes which can give rise to a similar topology are
$Z(\gamma)$ with $Z\to\qq$, $WW$ events with at least one $W$ decaying 
into hadrons, other four-fermion final states from neutral current
processes\footnote{The definition of neutral current (NC) and
charged current (CC) four-fermion processes of Ref.~\cite{CCNC} are
used throughout the text.} ($l^+l^-\qq$, 
$\nu\bar{\nu}\qq$, the latter being topologically identical to the signal) 
and events induced by two-photon collisions, hereafter called
two-photon events.    
Some of these processes have cross-sections larger than that of the
signal by several orders of magnitude. 
Therefore the analysis was performed in two steps: a preselection 
of hadronic events and a final selection based on a  
Feed-Forward Artificial Neural Network~\cite{jetnet}.

A sample of hadronic events was preselected by requiring at least seven
charged particles to be measured in the detector. 
Events from Bhabha scattering were rejected by a  cut on the total
electromagnetic 
energy, $E_{EM}/\sqrt{s} < 50\%$. The contribution from
two-photon collisions was reduced by requiring the total visible
energy to be larger than 20\% of $\sqrt{s}$ and the total transverse
energy, computed as the sum of the moduli of the momenta of each particle 
projected in the plane transverse to the beam axis, to be at least
15\% of $\sqrt{s}$.   
In addition, it was required that the half-opening angle of
the cone around the beam axis containing 15\% of the visible
energy had to be larger than 10$\deg$; two-photon collision events are
concentrated in the forward regions and have low values of this
variable. 
The background from $\epem\to \qq(\gamma)$ was reduced by
requiring the 
polar angle of the missing
momentum to satisfy the condition $|\cos\theta_{miss}|<0.98$ and the
acoplanarity\footnote{Acoplanarity is defined as the supplement of the
  angle between the projections of the two hadronic jets in the plane
  transverse to the beam direction.}  to be larger than 10$\deg$.  
$Z(\gamma)$ events, with $Z\to\qq$, were further suppressed by vetoing
events with electromagnetic clusters with energy larger than 45 GeV
or, if the
ISR photon escaped
undetected in the dead region between the barrel and end-cap
electromagnetic calorimeters ($\theta\sim 40\deg$),
by vetoing events with signals in the hermeticity counters
in a cone of 30$\deg$ around the direction of the missing momentum.
Selections on the maximum total multiplicity 
of charged and neutral tracks ( $<50$ ) and on
the visible mass 
(between 30 and 100 GeV/$c^2$) were applied to suppress the residual
contamination of multi-jet events from $WW$ or NC processes.
Finally, $WW$ events with one $W$ decaying to leptons were suppressed
by requiring no identified electron or muon with energy larger than
10\% or 7.5\% of $\sqrt{s}$, respectively. 
Particles were identified as muons if there was at least 
one muon chamber hit associated to a track or if the size and
longitudinal profile of the energy deposits associated to a track
in the hadronic calorimeters were consistent with a minimum ionizing particle.
Electron identification was based on the reconstructed showers in the
electromagnetic calorimeters associated to charged particle tracks.

The expected composition of the residual sample after the preselection
stage is shown in Table~\ref{tab:presel}, together with the number of
selected events at each centre-of-mass energy. At this level of the
selection, the fraction of signal events is between 6\% and 10\% at all the
energy points.

\begin{table}[tb]
\begin{center}
\vspace{0.5cm}
\begin{tabular}{|c|c|c|c|c|c|c|}
\hline\hline
 $\sqrt{s}$ (GeV) & $e^-\bar{\nu}_e q\bar{q}'$ signal & Other CC & NC & $q\bar{q}(\gamma)$ 
 & Total MC & Data \\
\hline\hline
 183 GeV & ~9.5 & ~66.8 & ~5.0 & ~74.2 &  $157.0\pm 0.7$ & 167 \\
 189 GeV & 32.5 & 195.6 & 25.7 & 204.5 &  $462.8\pm 2.0$ & 467 \\ 
 192-202 GeV & 54.2 & 276.9 & 51.4 & 227.3 &  $615.4\pm 1.7$ & 675 \\ 
 205-207 GeV & 55.6 & 240.5 & 54.4 & 194.9 &  $549.2\pm 1.7$ & 569 \\ 
\hline \hline
\end{tabular}
\end{center}
\caption{Number of events expected from the contribution of different
       channels and observed in the data after preselection of 
       $e^-\bar{\nu}_e q\bar{q}'$ events for the different
       years of data taking. ``Other CC'' indicates charged current
       processes different from the signal, ``NC'' indicates neutral current
       processes. Other final states, mainly  $\gamma\gamma$, contribute 
       less than 1\% of the total number of events. 
       The quoted errors on the total number of expected
       events (``Total MC'') are the ones due to limited Monte Carlo
       statistics. }
\label{tab:presel}
\end{table}
The final selection of $e^-\bar{\nu}_e q\bar{q}'$ events was based on a
Neural Network analysis. The input variables were chosen to provide a
good separation from the main residual backgrounds after preselection.
The first set of variables, listed below, discriminated the signal from  
$q\bar{q}(\gamma)$ events:
\begin{itemize}
\item effective centre-of-mass energy after ISR, $\sqrt{s'}$, 
      normalised to the centre-of-mass energy, $\sqrt{s}$~\cite{sprime}; 
\item normalised sum of the particle momenta projected on to the thrust axis,
      $P_l^{tot}/\sqrt{s}$;
\item cosine of the polar angle of the missing momentum, $|\cos\theta_{miss}|$;
\item normalised total missing momentum, $P_{miss}^{tot}/\sqrt{s}$; 
\item normalised total transverse momentum with
      respect to the beam axis,  
      $P_t^{tot}/\sqrt{s}$;
\item event thrust;
\item $|90\deg - \theta_{thrust}|$, where $\theta_{thrust}$ is the polar
  angle of the thrust axis. 
\end{itemize}
A second set of variables suppressed  $\tau\nu_\tau q\bar{q}'$ events, 
where the $\tau$ lepton produces an isolated particle or a low
multiplicity jet, and $\nu\bar{\nu}q\bar{q}$ events, where the kinematic
properties of the visible system should be consistent with the decay of
a $Z$:
\begin{itemize}
\item maximum transverse momentum of any particle with respect to
      the nearest jet, $P_{tJ}^{max}$, when the particles are clustered
      using the LUCLUS~\cite{luclus} algorithm with the parameter
      $d_{min}=6.5$ GeV/$c$; 
\item invariant mass of the detected particles in the event rescaled
      by the ratio of the total centre-of-mass energy to the visible
      energy, $M_{vis}\cdot\sqrt{s}/E_{vis}$;  
\item Lorentz boost factor of the visible part of the
      event in the laboratory frame,
      $\beta=P^{tot}/E_{vis}$. 
\end{itemize}

\begin{figure}[p]
\begin{center}
\epsfig{file=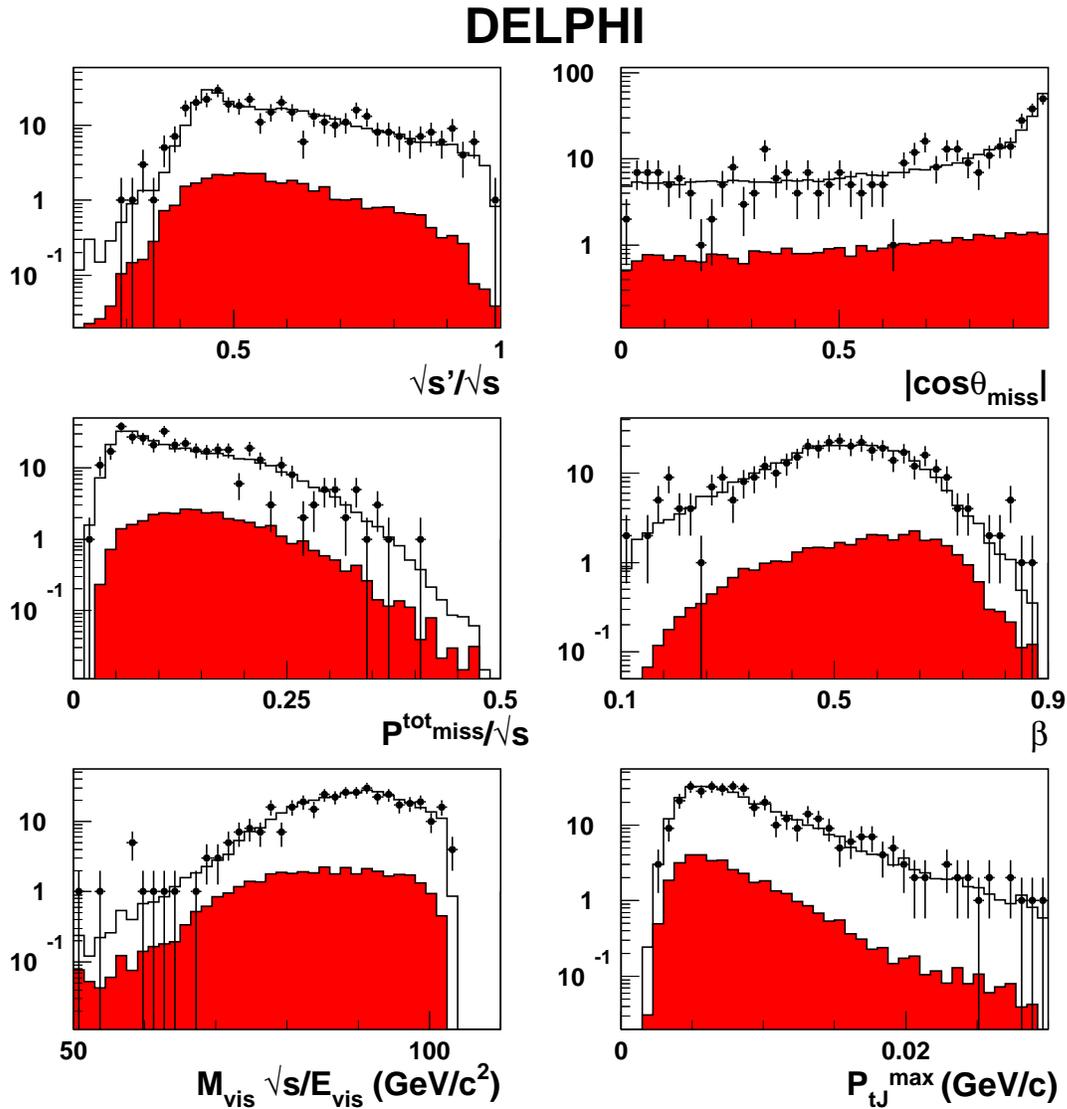,height=16cm}
\end{center}
\caption{$\wev$ channel $(W \to q\bar{q}')$ at $\sqrt{s}=200$~GeV. 
  Distribution of some Neural Network input variables,
  as defined in the text,
  in the real data (points with error bars) and in the
  simulation of the Standard Model predictions (histograms) after the preselection stage (see text). The
  distributions of these variables for the $\wev$ signal are shown as
  well (filled histograms). The $|\cos\theta_{miss}|$ distribution (top-right) stops 
  at 0.98, as explained in the text.
   }
\label{fig:nninp}
\end{figure}

\begin{figure}[bh]
\begin{center}
\epsfig{file=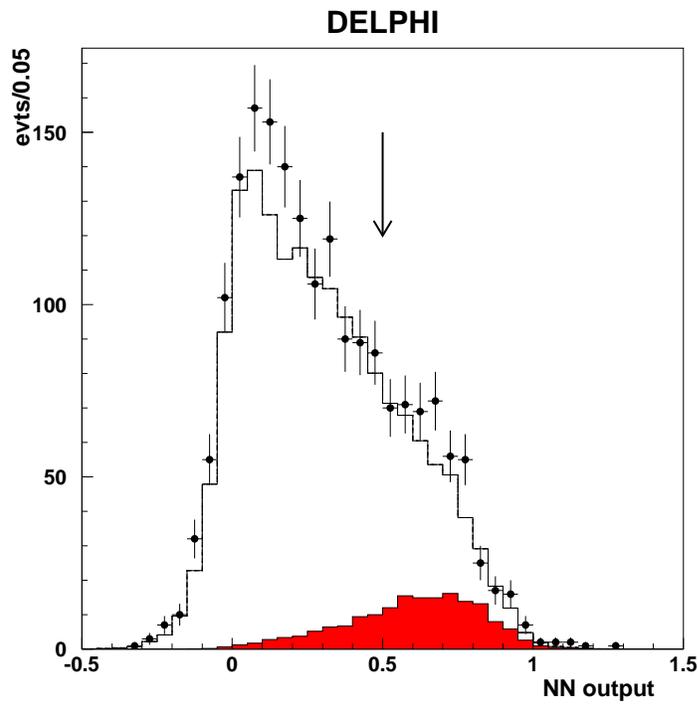,height=10cm}
\end{center}
\caption{$\wev$ channel $(W \to q\bar{q}')$ summed over all centre-of-mass energies:
  distribution of the Neural Network output variable
  in the real data (points with error bars) and in the
  simulation of the Standard Model predictions (histograms). The filled histogram represents the
  single-$W$ signal, the open area is the background expectation. 
  The arrow indicates the cut applied on this variable for the final event
selection. }
\label{fig:nnout}
\end{figure}

Distributions of some of these variables, at $\sqrt{s}=200$ GeV,
are shown in Figure~\ref{fig:nninp}.
The Neural Network was trained on samples of simulated events
including the signal, for which the output value was set to 1, and
the main backgrounds, $q\bar{q}(\gamma)$ and four-fermion processes, with
output set to 0. 
The distribution of the Neural Network output variable is shown in
Figure~\ref{fig:nnout}. The whole data sample is included in the
plot. The cut on the output variable was set at 0.5, 
the value for which the product of efficiency and purity was found to
be maximum.
In the region $50~{\mathrm GeV}/c^2 <M_{vis}\cdot\sqrt{s}/E_{vis} <
60~{\mathrm GeV}/c^2$ an excess of real data was found at each energy
point in the final selected sample (7.4$\pm$1.3\% of the total number
of events in the real data compared to 4.2$\pm$0.2\% in the
simulation). 
To account for this excess, a correction of 34 fb was added to
the background cross-section at each energy point.

The efficiency of the selection for the signal, the expected background, 
the luminosity and the number of selected events in the data at the
various centre-of-mass energies are reported in Table~\ref{tab:sigqq},
together with the cross-section for the hadronic channel alone,
evaluated from a fit of Poisson probabilities to the observed numbers
of events.   

Efficiencies and backgrounds at $\sqrt{s}=207$ GeV were found
to be compatible in the two periods with the sector S6 of the TPC on
or off, and results have been merged in the table.
The purity of the final selected sample is between 25\% and 30\%. 
The main contamination is due to $WW\to \tau\nu_\tau q\bar{q}'$ events.  

%
%
\begin{table}[tb]
\begin{center}
\vspace{0.5cm}
\begin{tabular}{|c|c|c|c|c|c|}
\hline\hline
$\sqrt{s}$ &  $\varepsilon$ & $\sigma_{bkg}$ & $\lu$ & $N_{data}$  &  $\sigma_{e\nu qq'}$  \\ 
(GeV) & (\%) & (pb) & (pb$^{-1}$) &  & (pb) \\ 
\hline \hline
 183 & 36.3$\pm$1.4 & 0.504$\pm$0.010 & ~51.6  & ~28 & $0.107^{+0.300}_{-0.107}$ \\
 189 & 37.0$\pm$1.5 & 0.506$\pm$0.011 & 153.8  & 110 & $0.565^{+0.190}_{-0.179}$ \\
 192 & 36.7$\pm$1.0 & 0.502$\pm$0.011 & ~24.5  & ~15 & $0.300^{+0.469}_{-0.300}$ \\
 196 & 35.2$\pm$0.6 & 0.504$\pm$0.009 & ~72.0  & ~49 & $0.502^{+0.290}_{-0.263}$ \\
 200 & 36.1$\pm$0.6 & 0.502$\pm$0.007 & ~81.8  & ~58 & $0.574^{+0.269}_{-0.247}$ \\
 202 & 37.5$\pm$1.0 & 0.503$\pm$0.010 & ~39.7  & ~30 & $0.674^{+0.391}_{-0.346}$ \\
 205 & 38.3$\pm$1.6 & 0.556$\pm$0.009 & ~66.2  & ~62 & $0.994^{+0.324}_{-0.298}$ \\
 207 & 39.2$\pm$1.5 & 0.560$\pm$0.010 & 129.7  & 114 & $0.814^{+0.217}_{-0.204}$ \\
\hline \hline
\end{tabular}
\end{center}
\caption{Performance (signal efficiency, $\varepsilon$, background cross-section, $\sigma_{bkg}$, 
  integrated luminosity, $\lu$, and number of selected events, $N_{data}$)
 of the $e^-\bar{\nu}_e q\bar{q}'$ event selection and measured
 cross-sections at the
 centre-of-mass energies considered in the analysis.
 The errors quoted on efficiencies and backgrounds are the ones due to
 limited Monte Carlo statistics.}
\label{tab:sigqq}
\end {table}

\subsection{Selection of leptonic events}
\label{sec:Wev_lept}

The experimental signature of the leptonic channel, 
$e^+e^-\to e^- \bar{\nu}_e l^+ \nu_l$, is the presence of a 
high energy lepton accompanied by a large missing 
momentum and no other significant energy deposition in the detector. 
The analysis was optimised for final state leptons that
are electrons or muons.
In both channels,
the contribution from $\evtv$ events was considered as part of the background.

The main backgrounds for the leptonic channel are the radiative
production of two leptons, $\eell(\gamma)$, $\eeww$ events and two-photon
collisions. 

Events were selected if exactly one well measured charged particle was 
reconstructed. The quality of the track measurement was assessed as
follows: 
\begin{itemize}
\item relative error on the momentum, $\Delta{p}/p$, smaller than $100\%$;
\item track length greater than 20~cm;
\item polar angle $\theta$ between $10\deg$ and $170\deg$;
\item impact parameter in the transverse plane, $|IP_{R\phi}|$, smaller
  than 4~cm, and that along the beam direction, $|IP_z|$, 
  smaller than 3~cm~$/\sin\theta$.
\end{itemize}
Loose identification criteria were applied, requiring associated hits in the
muon chambers or a significant energy deposit in the
electromagnetic calorimeter.
For electrons, the acceptance was
restricted to the barrel region, $|\cos\theta|<0.72$, and the best
determination of the electron energy was estimated by combining the
momentum measurement from the tracking devices and the 
measurement of the energy deposited in the calorimeters. 
Any other energy deposit in the detector unassociated to the lepton candidate
was required not to exceed 2~GeV. In addition, the
presence of tracks not fulfilling the quality criteria
listed above was used to veto the event.
The acceptance was restricted to the kinematic region of $W$ decays by
requiring the lepton momentum to lie below 45\% of
$\sqrt{s}$ and its transverse momentum to exceed 12\% of $\sqrt{s}$.

A large residual contamination was still present, due to cosmic-ray events
in the muon channel and to Compton scattering of a radiated photon in the 
electron channel. The former were suppressed by tightening the
selections on the track impact parameters to $|IP_{R\phi}|<0.2$~cm and
$|IP_z|<2$~cm for the muons. 
Compton scattering can mimic the $W^+\to{e^+}\nu_e$ signal when
the photon balancing the electron in the transverse plane is lost in
the dead region between the barrel and forward electromagnetic
calorimeters. Therefore events were rejected if a signal was found in
the hermeticity counters at an azimuthal angle larger than 90$\deg$ 
from the electron.

\begin{figure}[p]
\begin{center}
 \(
  \begin{array}{c} 
  \scalebox{0.5}{\epsfig{file=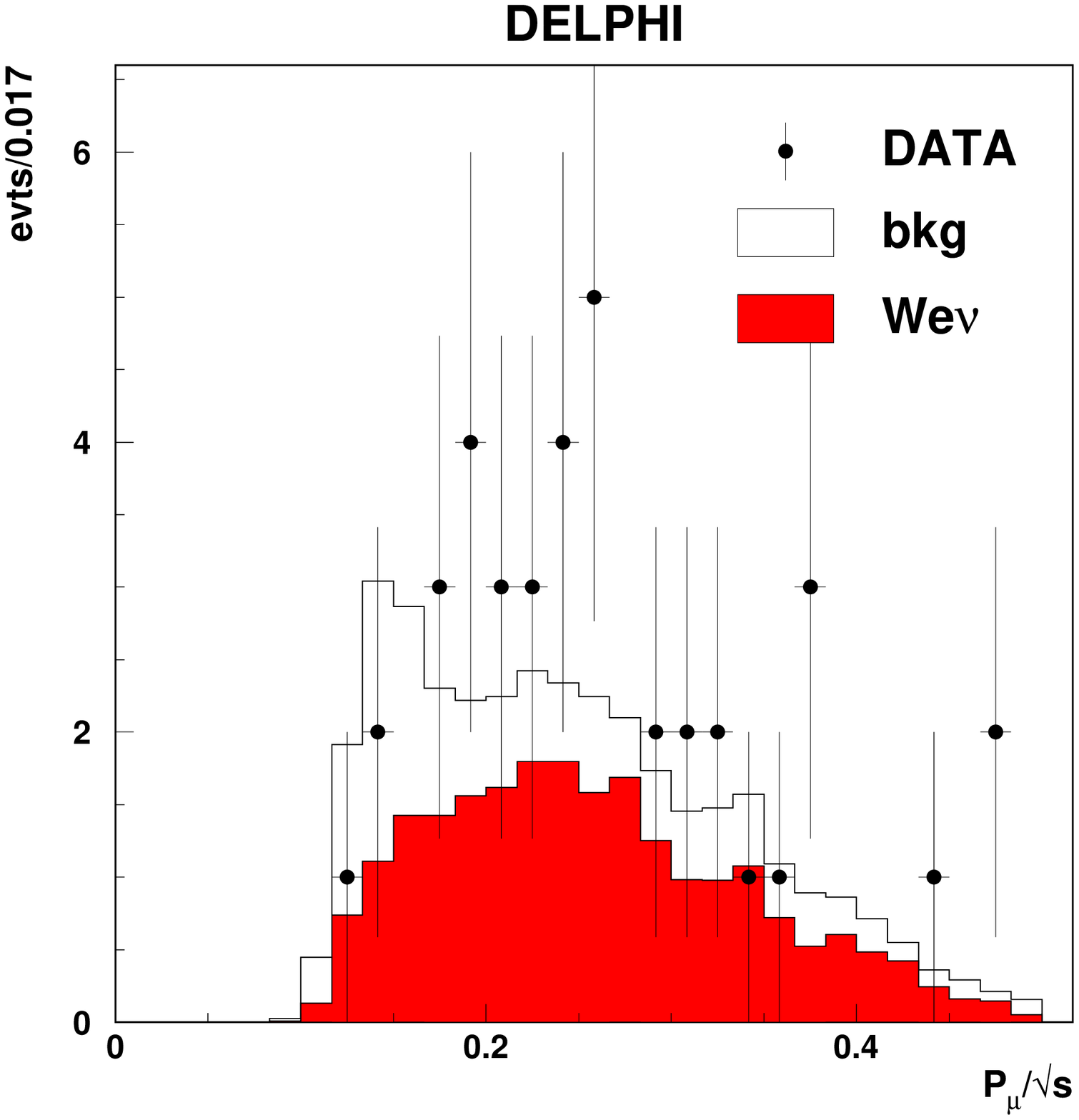}}\\
  \scalebox{0.5}{\epsfig{file=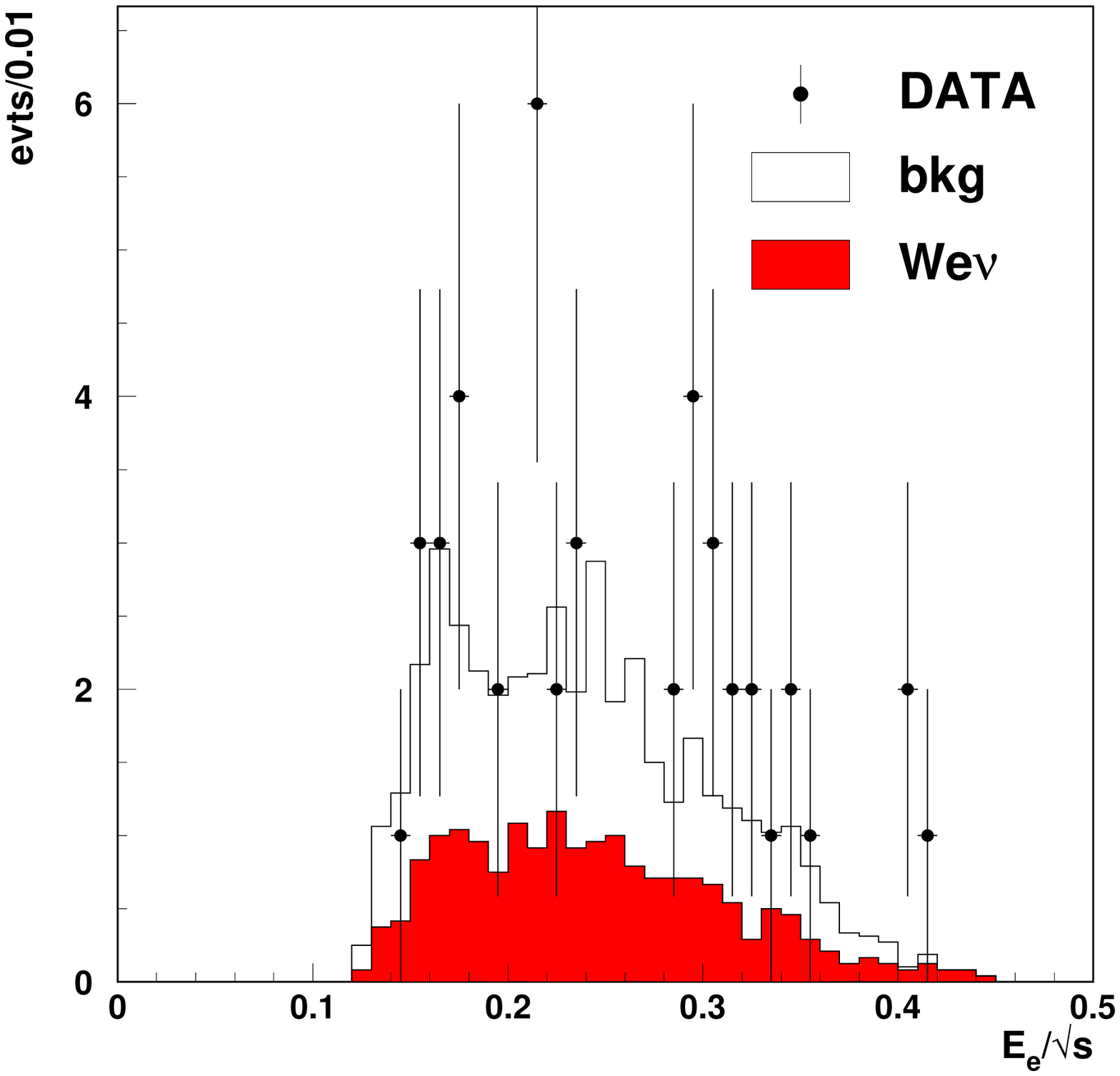}}
  \end{array}
 \)
\end{center}
\caption{$\wev$ channel $(W \to  l^+ \nu_l)$ summed over all centre-of-mass energies: 
  momentum distribution of the muon {\it (top)} and energy
  distribution of the electron {\it (bottom)} 
  in the real data (points with error bars) and in the simulation of the Standard Model predictions
  (histograms) for the events selected at the end of the analysis. The
  filled histograms represent the single-$W$ signal, the open area is
  the background expectation. }
\label{fig:plept}
\end{figure}

The distributions in data and simulation of the momentum of selected single
muons and of the energy of selected single electrons
are shown in Figure~\ref{fig:plept}.

The performance of the analysis at the various centre-of-mass 
energy values and the results obtained are reported in
Tables~\ref{tab:evmv} and~\ref{tab:evev}, respectively, for 
the $e^- \bar{\nu}_e \mu^+ \nu_\mu$ and $e^- \bar{\nu}_e e^+ \nu_e$ channels.
Due to the low statistics of selected events only the upper limits of
cross-sections at 95\% C.L. are given for each individual energy point.
The limits were derived following a Bayesian approach from the integration
of the Poissonian probabilities constructed with the number of
events selected in the data and predicted in the simulation.
For the electron channel a difference was found for efficiencies and
backgrounds corresponding to the two periods at $\sqrt{s}=207$~GeV
with TPC sector S6 on or off, and in Table~\ref{tab:evev} the 
weighted averages of the two are shown. Compatible values were found for the muon channel.

\begin{table}[tb]
\begin{center}
\vspace{0.5cm}
\begin{tabular}{|c|c|c|c|c|c|}
\hline\hline
$\sqrt{s}$  &$\varepsilon$ &   
$\sigma_{bkg}$ & $\lu$ & $N_{data}$ & $\sigma_{e\nu\mu\nu}$  \\
(GeV) & (\%)&   (fb) & (fb$^{-1}$) & & (fb) \\
\hline \hline
 183 & 44.8$\pm$2.8 & 18.8$\pm$1.6 & 0.0516  & ~7 & $<526$ at 95\% C.L. \\  
 189 & 47.2$\pm$1.7 & 19.1$\pm$1.2 & 0.1538  & ~5 & $<106$ at 95\% C.L. \\  
 192 & 48.4$\pm$2.7 & 18.6$\pm$1.6 & 0.0245  & ~1 & $<369$ at 95\% C.L. \\  
 196 & 49.0$\pm$1.6 & 20.2$\pm$1.3 & 0.0720  & ~4 & $<218$ at 95\% C.L. \\  
 200 & 45.2$\pm$2.5 & 22.8$\pm$1.4 & 0.0818  & ~7 & $<304$ at 95\% C.L. \\  
 202 & 45.3$\pm$1.7 & 24.0$\pm$2.0 & 0.0397  & ~5 & $<531$ at 95\% C.L. \\
 205 & 45.4$\pm$1.7 & 20.3$\pm$1.7 & 0.0662  & ~2 & $<172$ at 95\% C.L. \\
 207 & 46.3$\pm$1.8 & 23.0$\pm$1.6 & 0.1297  & ~8 & $<191$ at 95\% C.L. \\
\hline\hline
\end{tabular}
\end{center}
\caption{Performance (signal efficiency, $\varepsilon$, background cross-section, $\sigma_{bkg}$, 
  integrated luminosity, $\lu$, and number of selected events, $N_{data}$) of the 
  $e^-\bar{\nu}_e\mu^+\nu_{\mu}$  
  event selection at the centre-of-mass energies considered in
  the analysis. 
 The errors quoted on efficiencies and backgrounds are the ones due to
 limited Monte Carlo statistics.} 
\label{tab:evmv}
\end {table}
\begin{table}[tb]
\begin{center}
\vspace{0.5cm}
\begin{tabular}{|c|c|c|c|c|c|}
\hline\hline
$\sqrt{s}$ &$\varepsilon$ &   
$\sigma_{bkg}$ & $\lu$  & $N_{data}$ & $\sigma_{e\nu e\nu}$  \\
(GeV) & (\%) & (fb) & (fb$^{-1}$) &  & (fb)  \\
\hline \hline
 183 & 37.2$\pm$3.3 & 36.4$\pm$2.5 & 0.0516  & ~3 & $<316$ at 95\% C.L. \\  
 189 & 35.6$\pm$2.1 & 38.6$\pm$2.5 & 0.1538  & 13 & $<268$ at 95\% C.L. \\  
 192 & 35.6$\pm$2.1 & 43.1$\pm$2.5 & 0.0245  & ~1 & $<468$ at 95\% C.L. \\  
 196 & 35.5$\pm$2.1 & 44.4$\pm$2.5 & 0.0720  & ~4 & $<248$ at 95\% C.L. \\  
 200 & 32.3$\pm$1.9 & 41.1$\pm$2.5 & 0.0818  & ~6 & $<324$ at 95\% C.L. \\  
 202 & 31.0$\pm$2.0 & 40.9$\pm$2.5 & 0.0397  & ~3 & $<508$ at 95\% C.L. \\
 205 & 29.6$\pm$3.0 & 38.3$\pm$2.6 & 0.0662  & ~3 & $<287$ at 95\% C.L. \\
 207 & 29.0$\pm$2.1 & 38.4$\pm$2.9 & 0.1297  & 11 & $<352$ at 95\% C.L. \\
\hline\hline
\end{tabular}
\end{center}
\caption{Performance (signal efficiency, $\varepsilon$, background cross-section, $\sigma_{bkg}$,
 integrated luminosity, $\lu$, and number of selected events, $N_{data}$) of the 
 $e^-\bar{\nu_e}e^+\nu_e$ event selection at the centre-of-mass energies 
 considered in the analysis. 
 The errors quoted on efficiencies and backgrounds are the ones due to
 limited Monte Carlo statistics.}
\label{tab:evev}
\end {table}
%
%
\subsection{Study of systematic uncertainties}

The main source of systematic uncertainty in the present measurement
is the knowledge of the background level in the selected samples. In
particular, as can be seen from Table~\ref{tab:presel}, in the
hadronic channel selection there is an excess of data of (5$\pm$2)\%
with respect to the expectation at the preselection stage, at which
the sample consists mainly of background events. 
Rescaling the background at the final stage of the selection by this factor 
leads to an average decrease of 69$\pm$28 fb of the measured
cross-section. This was considered as a systematic error fully correlated
between the energy points. 
The correction factor added to the background cross-section,
accounting for the excess of real data in the region
$50~{\mathrm GeV}/c^2 <M_{vis}\cdot\sqrt{s}/E_{vis}< 60~{\mathrm GeV}/c^2$,
was known with a statistical uncertainty of $\pm10$ fb. This leads to 
a systematic error on $\sigma_{e\nu qq'}$ of $\pm26$ fb, fully
correlated between the energy points. 
  
Possible inaccuracies in the modelling of background processes were
evaluated by comparing different Monte Carlo generators. The only
effect was found in the $\qq(\gamma)$ channel: using the
ARIADNE~\cite{ariadne} event generator instead of PYTHIA,  
the background estimate was 508$\pm$14 fb instead of 502$\pm$7 fb
in the final $e^-\bar{\nu}_e q\bar{q}'$ sample selected at 200 GeV. 
The largest of the statistical errors of the ARIADNE and PYTHIA samples
was taken as a systematic error.
The total systematic error on the background cross-section,
due to the modelling of $\qq(\gamma)$ process and to the
limited simulation statistics (see Table~\ref{tab:sigqq}), amounts
approximately to $\pm3\%$ in the hadronic channel.

In the leptonic channels the systematic error on the background
cross-section due to the limited simulation statistics (see
Tables~\ref{tab:evmv} and~\ref{tab:evev}) amounts approximately to $\pm6\%$. 

From a comparison of dimuon events in data and simulation, the
tracking efficiency, $\varepsilon_{track}$,
of {\mbox DELPHI} was found to be 0.5\% higher in the simulation. This
value was assumed as a systematic error. 
This has a negligible effect on the background, while it affects the
selection efficiency of the signal for leptonic decays of the $W$.

The uncertainty on the efficiency of the electron identification,
$\varepsilon_e$, was 
estimated by comparing a sample of Bhabha events in data and
simulation. The discrepancy was at the level of 2\%.
The trigger efficiency for the leptonic events is known with an error
better than 1\%. This leads to a systematic error negligible compared
to the other sources considered.  

The luminosity is known with a total relative error of $\pm0.6\%$. 

\subsection{Total single-$W$ cross-section}

The total single-$W$ cross-section is defined as: 
\begin{equation}
\sigma_{e\nu ff'} = \sigma_{e\nu qq'} + 2 \times \sigma_{e\nu\mu\nu} + \sigma_{e\nu e\nu},
\label{eq:evff}
\end{equation}
where the factor two accounts for the  $e^-\bar{\nu}_e\tau^+\nu_{\tau}$ channel,
not measured in the present analysis, assuming $\mu-\tau$ universality.
This assumption introduces a theoretical error at the level of $\sim 3
\%$ on the  $e^-\bar{\nu}_e\tau^+\nu_{\tau}$ estimation. 

The effects of the uncertainties listed in the previous section on the
measurement of the 
$e^-\bar{\nu}_e f\bar{f}'$ cross-section at $\sqrt{s}$=200~GeV are given in 
Table~\ref{tab:syst}. The total systematic error, obtained from the
sum in quadrature of the individual contributions, 
is at the level of  $\pm9\%$.
For the measurement at the other centre-of-mass energies, the same
relative error was assumed. 

\begin{table}[tb]
\begin{center}
\vspace{0.5cm}
\begin{tabular}{|l|c|c|c|}
\hline\hline
 Systematic effect  &   Error on  & Error on  & Error on  \\
  ~          &  $\sigma_{e\nu qq'}$ (pb)  &
            $\sigma_{e\nu\mu\nu}$ (pb) & $\sigma_{e\nu e\nu}$ (pb) \\ \hline\hline
 $\Delta\sigma_{bkg}$ ($\evqq$) from preselection &   0.069  &    -    & - \\
 $\Delta\sigma_{bkg}$ ($\evqq$) from $M_{vis}\cdot\sqrt{s}/E_{vis}$ & 0.026 & - & - \\
 $\Delta\sigma_{bkg}$ ($\evqq$) $\pm 3\%$      &   0.042  &    -    & - \\
 $\Delta\sigma_{bkg}$ ($\evmv$) $\pm 6\%$      &   -      &  0.003 &    -   \\
 $\Delta\sigma_{bkg}$ ($\evev$) $\pm 6\%$      &    -     &    -    & 0.008 \\
\hline
 $\Delta\varepsilon$ ($\evqq$) due to simulation statistics   &   0.010 &   -   &  -   \\
 $\Delta\varepsilon$ ($\evlv$) due to simulation statistics   &   -    & 0.007  & 0.006 \\
\hline
 $\Delta\varepsilon$ ($\evlv$) due to $\varepsilon_{track}$
                                              & -       &  0.002 & 0.002\\
 $\Delta\varepsilon$ ($\evev$) due to $\varepsilon_e$
                                              & -       &     -   & 0.006 \\
\hline
 Luminosity  $\pm 0.6\%$                      &   0.012 &  0.001  & 0.001 \\
 \hline\hline
 Total                                        &   0.086  & 0.008  & 0.012 \\
 \hline\hline
\end{tabular}
\end{center}
\caption{Contributions to the systematic uncertainty
on the $e^-\bar{\nu}_e q\bar{q}'$, $e^-\bar{\nu}_e\mu^+\nu_{\mu}$ and $e^-\bar{\nu_e}e^+\nu_e$ cross-sections at
$\sqrt{s}=200$~GeV. The selection efficiency, $\varepsilon$, is as
defined in tables~\ref{tab:sigqq}, \ref{tab:evmv} and \ref{tab:evev};
$\varepsilon_{track}$ is the overall tracking efficiency in DELPHI,
and $\varepsilon_{e}$ is the electron identification efficiency.}
\label{tab:syst}
\end {table}

The values of $\sigma_{e\nu ff'}$ measured at the different centre-of-mass
energies together with their statistical and systematic errors are
shown in Table~\ref{tab:evff}. 

\begin{table}[htb]
\begin{center}
\vspace{0.5cm}
\begin{tabular}{|c|c|}
\hline\hline
$\sqrt{s}$ (GeV)  &  $\sigma_{e\nu ff'}$ (pb)      \\
\hline \hline
 183  &  $0.69^{+0.41}_{-0.23} \pm 0.06$  \\
 189  &  $0.75^{+0.22}_{-0.20} \pm 0.07$  \\
 192  &  $0.39^{+0.54}_{-0.31} \pm 0.04$  \\
 196  &  $0.68^{+0.33}_{-0.28} \pm 0.06$  \\
 200  &  $0.96^{+0.33}_{-0.29} \pm 0.09$  \\
 202  &  $1.24^{+0.51}_{-0.42} \pm 0.11$  \\
 205  &  $1.06^{+0.36}_{-0.30} \pm 0.10$  \\
 207  &  $1.14^{+0.26}_{-0.24} \pm 0.10$  \\
\hline\hline
\end{tabular}
\end{center}
\caption{Total single-W cross-section, as defined in the text
  (eq.~\ref{eq:evff}), as measured at the different centre-of-mass
  energies considered in the analysis. The first error is statistical
and the second one is systematic.}
\label{tab:evff}
\end {table}
%
\nopagebreak
\section{Single-\mbox{\boldmath $Z$} analysis}\label{sec:singleZ}

In the single $\gs/Z$ analysis, decays of the vector boson into hadronic and
$\mm$ final states were considered. Both final states are
characterized by an electron scattered at large angle
with respect to the incoming direction.  
The other electron, lost in the beam pipe, results in a
missing momentum pointing along the beam direction.

\subsection{Selection of hadronic events}
\label{sec:ZeeQQ}
The experimental signature of these events consists of a pair of jets
produced in the hadronic decay of the $\gs/Z$ recoiling against an
electron. 
To maximize the sensitivity of the analysis in the widest possible range of 
invariant masses of the $\gs/Z$, the event selection was performed in three steps:
\begin{enumerate}
\item a loose preselection of events;
\item the identification of an isolated electron;
\item the final selection of signal events, optimized differently in
  two ranges of the invariant mass of the hadronic system, $m_{q\qbar}$,
  according to the most relevant background process in each region.
\end{enumerate}

\begin{table}[t]
\begin{center}      
\vspace{0.5cm}
\begin{tabular}{|l|ccccc|c|r|}
\hline\hline
       & $\gs/Zee$     & $WW$           &  $Z(\gamma)$ & $\gamma\gamma$ &  Others        &     Total MC  & Data
\\ \hline\hline                                                               
183 GeV & & & & & & & \\ \hline
Preselection                                                                   
      & 24.2  &  202.9   &  560.5  &  160.3  &  149.9  &  1097.8  &  1238          
\\ \hline                                                                     
$e$ ident.                                                            
      & 18.2  &  75.5    &   23.0  &   21.7  &   58.1   &  196.5   & 195     
\\ \hline                                                                     
Signal selection
       & 11.4$\pm$0.2  &  0.4$\pm$0.1  & 2.4$\pm$0.3 & 0.8$\pm$0.8  &  1.3$\pm$0.1   & 16.3$\pm$0.9 & 23         
\\ \hline\hline
189 GeV& & & & & & & \\ \hline
Preselection                                                                   
      & 73.5  &  647.4   & 1487.6  &  434.1  &  426.6  &  3069.2  &  3470          
\\ \hline                                                                     
$e$ ident.                                                            
      & 55.9  & 244.2    &   65.7  &   62.6  &  168.2   &  596.6   & 577     
\\ \hline                                                                     
Signal selection
       & 34.9$\pm$0.4  &  1.4$\pm$0.2  & 6.5$\pm$0.7 & 3.3$\pm$1.2  &  4.5$\pm$0.5   & 50.7$\pm$1.5 & 54         
\\ \hline\hline
192-202 GeV& & & & & & & \\ \hline
Preselection                                                                   
      &113.1  &  985.1   & 1946.4  &  669.4  &  608.1  &  4322.0  &  5016          
\\ \hline                                                                     
$e$ ident.                                                            
      & 85.4  & 382.9    &   87.0  &   70.5  &  237.3   &  863.1   & 915     
\\ \hline                                                                     
Signal selection
       & 54.8$\pm$0.5  &  2.7$\pm$0.2  & 9.4$\pm$0.6 & 2.6$\pm$0.7  &  6.0$\pm$0.3   & 75.6$\pm$1.0 & 78         
\\ \hline\hline
205-207 GeV& & & & & & & \\ \hline
Preselection                                                                   
      &110.1  &  938.2   & 1620.6  &  654.2  &  562.2  &  3885.3  &  4034          
\\ \hline                                                                     
$e$ ident.                                                            
      & 83.5  & 374.4    &   75.5  &   78.7  &  212.5   &  824.6   & 786     
\\ \hline                                                                     
Signal selection
       & 54.6$\pm$0.5  &  2.9$\pm$0.2  & 8.8$\pm$0.5 & 1.9$\pm$0.9  &  6.1$\pm$0.2   & 74.3$\pm$1.1 & 76         
\\ \hline\hline

\end{tabular}
\end{center}    
\caption{Number of events expected from the contributions of different
       channels and observed in the data at different stages of the
       $\gs/Zee$ selection (hadronic channel) for the different
       years of data taking.  
       The column labelled ``$\gamma\gamma$'' refers to resolved
       two-photon events. 
       The column labelled ``Others'' includes other four-fermion
       processes, namely $eeqq$ outside the signal definition and
       $\gs/Zee$ with fully leptonic final state, and events from
       Bhabha scattering, the four-fermion processes supplying the
       more important contribution.}
\label{tab:Zee1}
\end {table}
\noindent The preselection of events consisted of the following
requirements:
\begin{itemize}
\item at least five charged particles in the event with at least one 
  in the TPC
  with a measured transverse momentum larger than 2.5~GeV/$c$, in
  order to select hadronic events; 
\item the presence of at least one electron candidate selected by
 requiring an energy deposit in the electromagnetic calorimeters
 $E_e > 3$~GeV, with an associated charged particle with $|\cos
 \theta_e| < 0.985$, corresponding to the acceptance of 
 the DELPHI's forward and barrel electromagnetic calorimeters; 
\item in events with more than one reconstructed electromagnetic
  shower, the energy of the second most energetic shower was required
  to be less than $0.6 E_{beam}$. This condition was imposed in order
  to reject events from Bhabha scattering. 
\end{itemize}
\noindent The electron candidates were then retained if they satisfied the
following criteria: 
\begin{itemize}
\item in the barrel region ($42^\circ<\theta<138^\circ$), the track
  parameters were required to be consistent with
  those of the shower measured by the electromagnetic calorimeter, with
  the additional requirement, for showers with energy higher than
  30 GeV, that the energy deposited in the hadronic calorimeter did not
  exceed 10\% of that deposited in the electromagnetic calorimeter; 
\item in the forward region ($10^\circ<\theta<32^\circ$ and
  $148^\circ<\theta<170^\circ$), the energy of the shower,
  reconstructed by re-clustering the energy deposits compatible with a
  single electromagnetic shower, was required to be compatible with
  the momentum of exactly one track measured in DELPHI's vertex
  detector and very forward tracker, and with the momentum of no more
  than one track measured in the inner detector and TPC
  (see~\cite{DELPHI,VD}) for detailed description of these detectors); 
\item the angle, $\alpha_1$, of candidate tracks with respect to the closest charged particle
  with momentum $p > 0.5$~GeV/$c$ had to lie in the range $15\deg <
  \alpha_1 < 170\deg$, where the upper limit was imposed to reject events from Bhabha scattering
  left in the sample at this stage of the selection;
\item the angle, $\alpha_2$, of candidate tracks with respect to the second closest charged particle
  with momentum $p > 0.5$~GeV/$c$ was required to satisfy the
  condition $\alpha_2 > 40\deg$. 
\end{itemize}
Electrons from conversions or from decays were further suppressed
by requiring their impact parameters with respect to the primary
interaction vertex to be within the range $|IP_{R\phi}|<0.35$~cm in the transverse
plane and  $|IP_{z}|<1$~cm along the beam line.

\begin{figure}[p]
\begin{center}
  \(
  \begin{array}{c}
    \scalebox{0.5}{\epsfig{file=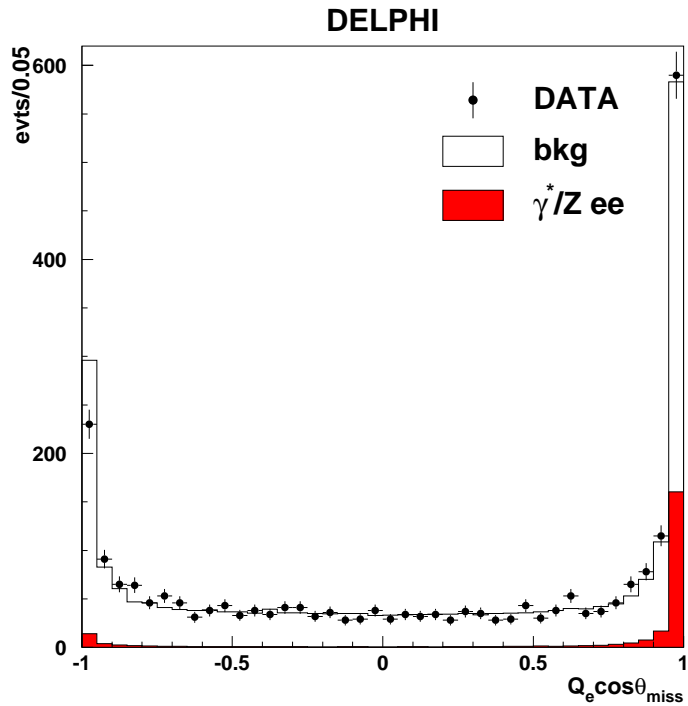}} \\
    \scalebox{0.5}{\epsfig{file=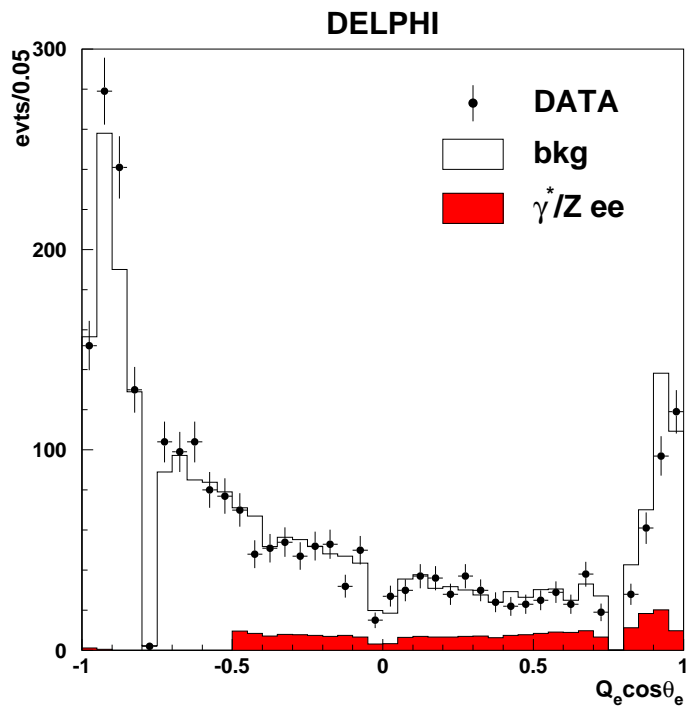}} 
  \end{array}
  \)
\end{center}
\caption{$\gs/Zee$ channel $(\gs/Z \to q\bar{q})$ summed over all centre-of-mass energies: 
  distributions of the variables used for the signal definition 
  at the reconstruction level
  after the ``electron identification'' step (see
  Section~\ref{sec:ZeeQQ}), in the real data
  (points with error bars) and in the simulation of the Standard Model predictions (histograms). The
  $\gs/Zee$ signal is defined in the kinematic region described in
  Section~\ref{sec:def}.} 
\label{fig:ZeeCuts}
\end{figure}

\begin{figure}[htb]
\begin{center}
  \(
  \begin{array}{c}
    \scalebox{0.5}{\epsfig{file=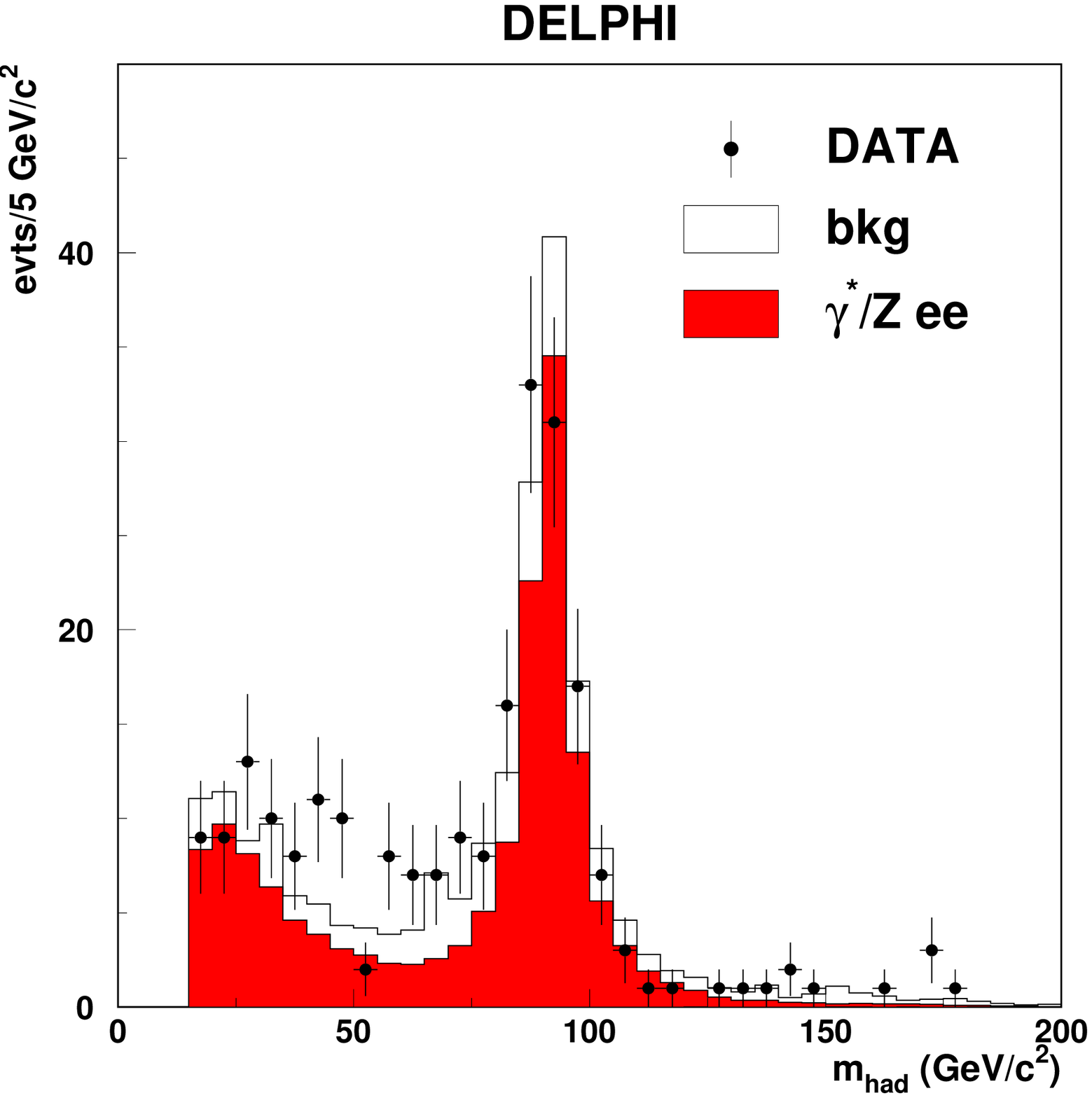}} \\
    \scalebox{0.5}{\epsfig{file=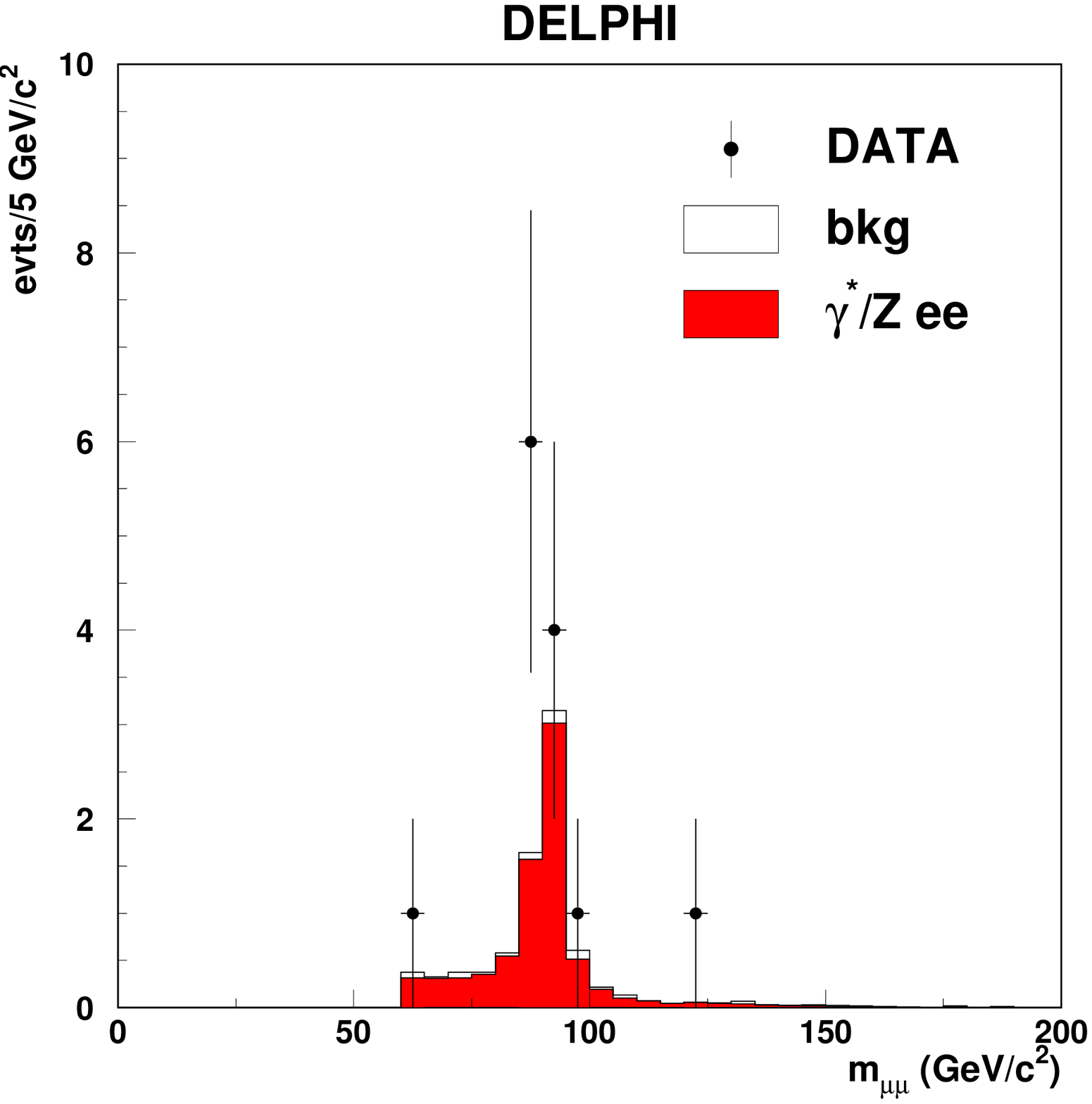}}
  \end{array}
  \)
\end{center}
\caption{$\gs/Zee$ channel summed over all centre-of-mass energies:
  invariant mass distribution of the $\gs/Z$ system 
  in the real data (points with error bars) and in the
  simulation of the Standard Model predictions (histograms) for hadronic {\it (top)}  and
  $\mm$ {\it (bottom)} final states, in the selected signal sample.}
\label{fig:ZeeMff}
\end{figure}

The charged and neutral particles were then clustered into two jets
with the Durham algorithm~\cite{durh}, excluding the tag electron.
Events for which the jet resolution variable that separates
the three-jet topology from the two-jet topology, $y_{3 \to 2}$, was
smaller than $10^{-4}$ were rejected. 
For the final selection a constrained kinematic fit of the event,
imposing energy and momentum conservation, was then performed assuming
a topology of signal events with two jets, a visible electron and one
lost along the beam line.  
The four-momentum of the invisible electron was chosen to be 
$(0,0,Q_{e}E,E)$ with $Q_{e}$ the charge of the tagged
electron.
Fits with a $\chi^2$ probability smaller than $10^{-5}$ were
rejected.
%
The final selection of signal events was then performed  
using the fitted kinematic variables for the tagged electron and the
hadronic system.
It was required that the following conditions be satisfied:
\begin{itemize}
\item $Q_{e} \cos \theta_{miss} > 0.95$, with $\theta_{miss}$ being
  the polar angle of the missing momentum computed
  before the kinematic fit;
\item $Q_{e} \cos \theta_{e} > -0.5$, with $\theta_e$ being the
  polar angle of the tagged electron.
\end{itemize}
The distributions of these variables after the electron identification
cuts are shown in Figure~\ref{fig:ZeeCuts} for the real and simulated
data. 
For $m_{q\qbar}<60$~GeV/$c^2$, where the dominant background consisted of
resolved $\gamma\gamma$ collisions, events with $Q_{e} \cos \theta_{e}
> 0.9$ and $E_{e}>0.75 E_{beam}$ were also rejected.
The numbers of selected data candidates and different background
contributions after each selection step
are shown in Table~\ref{tab:Zee1} for the different years of data-taking, while
Table~\ref{tab:Zee2} shows the composition of the entire sample after the final
selection in the two mass ranges.
An excess of data of about 10\% is observed at preselection level
mostly due to imperfectly simulated events from Bhabha scattering.
The efficiency of the selection on the signal, the expected background
and the number of selected events in the data at the eight centre-of-mass
energies are reported in Table~\ref{tab:ZeePerf}, together with the evaluated 
cross-section. \\ 
%
%
\begin{table}[t]
\begin{center}   
\vspace{0.5cm}
\begin{tabular}{|c|ccccc|c|r|}
\hline\hline
    mass range    & $\gs/Zee$     & $WW$           &  $Z(\gamma)$ &
    $\gamma\gamma$ &  Others        &     Total MC  & Data \\
(GeV/$c^2$) & & & & & & & 
\\ \hline\hline                                                               
$15 < m_{q\qbar} < 60$
       & 49.2$\pm$0.5  &  0.2$\pm$0.1  & 3.6$\pm$0.4 & 4.2$\pm$1.2  & 9.4$\pm$0.4   & 66.7$\pm$1.5 & 80         
\\ \hline                                                                     
$m_{q\qbar} > 60$ 
       &106.5$\pm$0.5  &  7.2$\pm$0.4  &23.5$\pm$1.0 & 4.3$\pm$1.3 & 8.5$\pm$0.2   &150.1$\pm$1.8 &151         
\\ \hline\hline
\end{tabular}
\end{center}   
\caption{Number of events expected from the contributions of different
       channels and observed in the data at the end of the 
       $\gs/Zee$ selection (hadronic channel) for the overall
       LEP2 sample, in the two invariant mass ranges.   
       The column labelled ``$\gamma\gamma$'' refers to resolved 
       two-photon events.
       The column labelled ``Others'' includes other four-fermion
       processes, namely $eeqq$ outside the signal definition and
       $\gs/Zee$ with fully leptonic final state, and events from
       Bhabha scattering, the four-fermion processes supplying the
       more important contribution.}
\label{tab:Zee2}
\end {table}
The distribution of the invariant mass of the hadronic
system after the
kinematic fit is shown in Figure~\ref{fig:ZeeMff} for the overall LEP2
sample.
The peak in the invariant mass distribution around the $Z$ mass 
corresponds to events for which the contribution of the $Zee$ process
is dominant.
\newlength{\LL} \settowidth{\LL}{$<201$ at 95\% C.L.}
\begin{table}[p]
\begin{center}
\vspace{0.5cm}
\begin{tabular}{|c|c|c|c|c|c|}
\multicolumn{6}{l}{$\gs/Z \to \qq~~~(15 < m_{q\qbar} < 60~{\mathrm GeV}/c^2)$}\\
\hline\hline
$\sqrt{s}$ & $\varepsilon$  &   $\sigma_{bkg}$ & $\lu$ &   $N_{data}$ &   $\sigma$ \\
 (GeV)     & (\%)  &  (pb)         & (pb$^{-1}$) & & (pb) \\
\hline \hline
 183     & 30.3$\pm$0.8 & 0.015$\pm$0.002 &  52.0 & 11 &$0.65^{+0.23}_{-0.19}\pm0.03$  \\
\hline                                                                                      
 189     & 30.7$\pm$0.8 & 0.030$\pm$0.006 & 153.5 & 16 &$0.24^{+0.09}_{-0.08}\pm0.02$ \\  
\hline                                                             
 192     & 32.1$\pm$0.8 & 0.027$\pm$0.006 &  25.1 &  6 &$0.66^{+0.35}_{-0.26}\pm0.04$ \\  
\hline                                                              
 196     & 29.9$\pm$0.8 & 0.021$\pm$0.004 &  75.9 & 14 &$0.55^{+0.18}_{-0.15}\pm0.03$ \\  
\hline                                                             
 200     & 29.4$\pm$0.8 & 0.026$\pm$0.005 &  82.8 &  6 &$0.16^{+0.12}_{-0.09}\pm0.02$ \\  
\hline                                                             
 202     & 29.0$\pm$0.8 & 0.026$\pm$0.005 &  40.3 &  2 &$0.08^{+0.15}_{-0.08}\pm0.03$ \\  
\hline                                                             
 205     & 29.8$\pm$0.7 & 0.021$\pm$0.004 &  75.9 & 12 &$0.46^{+0.17}_{-0.14}\pm0.03$ \\  
\hline 
 207     TPC OK     & 28.3$\pm$0.9 & 0.019$\pm$0.004 & 84.1 &  4 & \multirow{2}{\LL}{~~$0.25^{+0.10}_{-0.09}\pm0.02$} \\  
         TPC S6-off & 27.4$\pm$0.7 & 0.030$\pm$0.014 & 51.4 &  9 &  \\
\hline\hline

\multicolumn{6}{l}{$\gs/Z \to \qq~~~(m_{q\qbar} > 60~{\mathrm GeV}/c^2)$}\\
\hline \hline
$\sqrt{s}$ &  $\varepsilon$  &   $\sigma_{bkg}$ & $\lu$ &   $N_{data}$ &   $\sigma$ \\
 (GeV)     & (\%)  &  (pb)         & (pb$^{-1}$) & & (pb) \\
\hline\hline
 183    & 27.2$\pm$0.4 & 0.078$\pm$0.017 &   52.0  & 12 &$0.56^{+0.27}_{-0.22}\pm0.07$ \\  
\hline                                                                                                           
 189    & 27.8$\pm$0.4 & 0.068$\pm$0.007 &  153.5  & 38 &$0.64^{+0.15}_{-0.14}\pm0.04$ \\
\hline                                                                                     
 192    & 28.1$\pm$0.4 & 0.063$\pm$0.006 &   25.1  &  6 &$0.63^{+0.40}_{-0.30}\pm0.04$ \\
\hline                                                                                     
 196    & 28.8$\pm$0.3 & 0.060$\pm$0.006 &   75.9  & 19 &$0.66^{+0.21}_{-0.18}\pm0.04$ \\
\hline                                                                                     
 200    & 29.7$\pm$0.5 & 0.072$\pm$0.006 &   82.8  & 20 &$0.57^{+0.20}_{-0.17}\pm0.04$ \\
\hline                                                                                     
 202    & 30.5$\pm$0.4 & 0.066$\pm$0.006 &   40.3  &  5 &$0.19^{+0.21}_{-0.16}\pm0.02$ \\
\hline                                                                                     
 205    & 30.7$\pm$0.3 & 0.072$\pm$0.006 &   75.9  & 14 &$0.37^{+0.17}_{-0.15}\pm0.03$ \\
\hline 
 207    TPC OK     & 31.0$\pm$0.3 & 0.068$\pm$0.006 & 84.1  & 22 & \multirow{2}{\LL}{~~$0.69^{+0.15}_{-0.14}\pm0.03$} \\
        TPC S6-off & 29.4$\pm$0.3 & 0.060$\pm$0.004 & 51.4  & 15 & \\
\hline\hline
\multicolumn{6}{l}{$\gs/Z \to \mm~~~(m_{\mm} > 60~{\mathrm GeV}/c^2)$}\\
\hline\hline
$\sqrt{s}$ &  $\varepsilon$  &   $\sigma_{bkg}$ & $\lu$ &   $N_{data}$ &   $\sigma$ \\
 (GeV)     & (\%)  &  (fb)         & (fb$^{-1}$) & & (fb) \\
\hline\hline
 183   & 27.4$\pm$1.1  & 0.6$\pm$0.2 & 0.0540 & 1 & $<319$ at 95\% C.L.  \\  
\hline                                                                  
 189   & 26.2$\pm$1.0  & 1.1$\pm$0.4 & 0.1581 & 5 & $<250$ at 95\% C.L. \\ 
\hline                                                                 
 192   & 26.3$\pm$1.0 &  0.7$\pm$0.2 & 0.0258 & 0 & $<441$ at 95\% C.L. \\ 
\hline                                                                 
 196   & 26.7$\pm$1.0 &  1.3$\pm$0.3 & 0.0769 & 2 & $<301$ at 95\% C.L. \\ 
\hline                                                                 
 200   & 27.2$\pm$1.0 &  1.2$\pm$0.3 & 0.0843 & 1 & $<203$ at 95\% C.L. \\ 
\hline                                                                 
 202   & 26.7$\pm$1.0 &  0.9$\pm$0.2 & 0.0411 & 0 & $<273$ at 95\% C.L. \\ 
\hline                                                                 
 205   & 26.4$\pm$1.0 &  0.6$\pm$0.2 & 0.0767 & 1 & $<232$ at 95\% C.L. \\ 
\hline 
 207    TPC OK     & 26.1$\pm$1.0 & 1.1$\pm$0.3 & 0.0874  & 1 & \multirow{2}{\LL}{$<201$ at 95\% C.L.} \\
        TPC S6-off & 27.5$\pm$1.0 & 0.8$\pm$0.2 & 0.0544  & 2 & \\
\hline\hline
\end{tabular}
\end{center}
\caption{Performance (signal efficiency, $\varepsilon$, background cross-section, $\sigma_{bkg}$, 
  integrated luminosity, $\lu$, and number of selected events, $N_{data}$)
  of the $\gs/Zee$ event selection at the
  centre-of-mass energies considered in the analysis. The period with
  TPC sector 6 down is indicated as ``207 TPC-S6 off''.
  Cross-sections for $\gs/Z \to \mm$ are expressed in femtobarns due to smaller values.}
\label{tab:ZeePerf}
\end {table}

\subsection{Selection of leptonic events}

The search was restricted to events with $\gs/Z$ going 
into a $\mu^+\mu^-$ pair with invariant mass 
above 60 GeV/c$^2$. The general features are the same as
for the hadronic channel with jets replaced by muons.
Thus a three-track signature, of two high-momentum muons and one
$e^+$ or $e^-$, scattered at large angle, is expected in the detector.
The signal selection criteria on angular distributions were similar 
to those used in the hadronic channel.

In the preselection the event was required to have exactly three tracks
fulfilling the following criteria:
\begin{itemize}
\item fractional error on the momentum $\Delta p/p < 50 \%$;
\item impact parameter in the transverse plane  $|IP_{R\phi}| < 0.5$~cm 
         and along the beam direction $|IP_{z}| < 3$~cm;
\item at least one associated hit in the Vertex Detector.
\end{itemize}
The sum of the charges of the three particles was required to
be $\pm$1.
Possible photon conversions were removed according to the  
procedure described in~\cite{DELPHI} and requiring in addition
the minimum opening angle of any track pair to be larger than
$5^\circ$. 

Since the event topology is simple, the particle identification 
required at least two tracks to be identified as leptons
($\mu$ or $e$) and at least one of them to be a muon. 
For muon identification loose 
criteria were applied, as in the case of single-$W$ production 
(see Section~\ref{sec:Wev_lept}).
The flavour of the possible unidentified track was inferred from
partial information taking into account the combination of the charges of the 
observed particles.
In the case of $\mu^+x^-e^{\pm}$ or $x^+\mu^-e^{\pm}$, the unidentified track 
$x$ was treated as $\mu$. For $\mu^+\mu^-x^{\pm}$ the track $x$ was taken 
as $e^{\pm}$. 
Since the efficiency of the identification of the electrons was
smaller than for the muons, a majority of events with
the $\mu^+\mu^-$ pair detected and an unidentified electron was accepted this
way. This reduced dramatically the sensitivity of event selection,
and hence the loss of efficiency, to the efficiency of 
electron identification (less than 5\% drop of signal selection
efficiency was observed after forcing the electron track to be always
unidentified). Events with two tracks identified as electrons were rejected.  
At the preselection stage, the momentum of the $e^{\pm}$
candidate had to be greater than 2 GeV/$c$, and the invariant mass of the
$\mu^+\mu^-$ pair greater than 20 GeV/$c^2$. 

Due to the stringent cut on low multiplicity of the event, 
the data reduction factor was large.
For all energy points, 88 events were preselected and 
94.0$\pm$0.6 events were expected. 
At this stage most of the events came 
from other neutral current four-fermion processes 
with $e^+e^- \mu^+\mu^-$ in the final state but outside
the kinematic limits of the signal definition
(see Table~\ref{bgsplit}). The remaining contributions 
came mainly from the neutral current four-fermion processes 
($e^+e^- \to l_1^+l_1^-l_2^+l_2^-~$ excluding the $e^+e^- \mu^+\mu^-$ case), from
two-fermion processes ($e^+e^- \to \mu^+\mu^-(\gamma)$ and $e^+e^- \to
\tau^+\tau^-(\gamma)$) and a small fraction from $e^+e^- \to W^+W^-$.
\begin{table}[tb]
\begin{center}
\begin{tabular}
{|l|ccc|c|r|}
\hline
\hline
& $\gs/Zee$ & $(e^+ e^- \mu^+\mu^-)_{bkg}$ & Others & Total MC & Data \\
\hline \hline
183 GeV & & & & &\\
\hline \hline  
Preselection     & 0.87$\pm$0.03 & 6.3$\pm$0.1 & 0.6$\pm$0.1 & 7.8$\pm$0.1 & 4 \\ 
\hline
Final Selection     & 0.60$\pm$0.02 & 0.008$\pm$0.003 & 0.02$\pm$0.01 & 0.63$\pm$0.03 & 1 \\ 
\hline \hline 
189 GeV  & & & & &  \\
\hline \hline
Preselection     & 2.6$\pm$0.1  & 19.2$\pm$0.4 & 1.8$\pm$0.2 & 23.6$\pm$0.4 & 24 \\
\hline
Final Selection     & 1.7$\pm$0.1 & 0.03$\pm$0.01 & 0.14$\pm$0.06 & 1.91$\pm$0.09 & 5 \\ 
\hline \hline
192-202 GeV  & & & & & \\
\hline \hline
Preselection     & 3.8$\pm$0.1 & 26.3$\pm$0.3 & 2.4$\pm$0.1 & 32.5$\pm$0.3 & 21 \\
\hline
Final Selection     & 2.7$\pm$0.1 & 0.04$\pm$0.01 & 0.21$\pm$0.03 & 2.93$\pm$0.06 & 3 \\ 
\hline \hline
2000  205-207 GeV  & & & & & \\
\hline \hline
Preselection     & 3.7$\pm$0.1 & 23.8$\pm$0.3 & 2.3$\pm$0.1 & 29.8$\pm$0.3 & 39 \\
\hline
Final Selection     & 2.6$\pm$0.1 & 0.04$\pm$0.01 & 0.15$\pm$0.03 & 2.79$\pm$0.06 & 4 \\ 
\hline \hline
\end{tabular}
\end{center}
\caption{Number of events expected from the contributions of different
       channels and observed in the data at different stages of the
       $\gs/Zee$ selection (leptonic channel) for the different years
       of data taking. 
       The column labelled ``$(e^+ e^- \mu^+\mu^-)_{bkg}$'' shows the numbers 
       for the background events coming from all processes with $e^+
       e^- \mu^+\mu^-$ in the final state not fulfilling the signal
       definition criteria. All other background sources are collected
       inside the column labelled ``Others''.} 
\label{bgsplit}
\end{table}

A kinematic fit was performed before applying the final selection cuts 
to the data. An electron lost along the beam line and no
missing momentum in the transverse plane were assumed.
The invariant mass of the $\mu^+\mu^-$ pair was recalculated if the probability
of the kinematic fit was above 0.001. 
The high purity of this channel implies that events with
lower fit probability still correspond to $e^+e^- \mu^+\mu^-$ production, but with a
poorly measured visible electron; such events were therefore retained
and the original uncorrected $\mu^+\mu^-$ invariant mass was kept. 
In agreement with the signal definition the $\mu^+\mu^-$ invariant 
mass was then required to be greater than 60 GeV/$c^2$,
the momentum of the observed electron greater than 3 GeV/$c$ and its polar angle
$\theta_e$ to satisfy the condition $Q_e \cos\theta_{e} >-0.5$, with
$Q_e$ being the charge of the observed electron. 

Finally the allowed angular ranges for the direction of the $Z/\gamma^*$
momentum and missing momentum were defined by the following conditions:
\begin{itemize}
\item
$Q_e \cos\theta_{\mu^+\mu^-} < -0.8$, with $\theta_{\mu^+\mu^-}$ 
being the polar angle of the $\mu^+\mu^-$ system;
\item
$Q_e \cos\theta_{miss} > 0.8$, with $\theta_{miss}$ being the polar angle 
of the missing momentum computed before the kinematic fit.
\end{itemize}
After the final selection the background contribution is expected to be less  
than 10\% of the total selected events. The remaining background from processes 
with $e^+ e^- \mu^+\mu^-$ in the final state which was dominant at the preselection level
was reduced to about 1\%.


The efficiency of the selection of the signal, the expected background
and the number of selected events in the data for all
centre-of-mass energies are reported in Table~\ref{tab:ZeePerf}.
Due to the low statistics of selected events only the upper limits of
cross-sections at 95\% C.L. are given for each individual energy point.
The limits were derived following a Bayesian approach from the integration
of the Poissonian probabilities constructed with the number of
events selected in the data and predicted in the simulation.
In total 13 events were selected and 8.3$\pm$0.1 events were expected 
from data in the energy range from 183 GeV to 207 GeV. 
The  $\mu^+\mu^-$ invariant mass distribution is shown 
in Figure~\ref{fig:ZeeMff}.

\begin{figure}[htb]
\begin{center}
  \(
  \begin{array}{c}
    \scalebox{0.5}{\epsfig{file=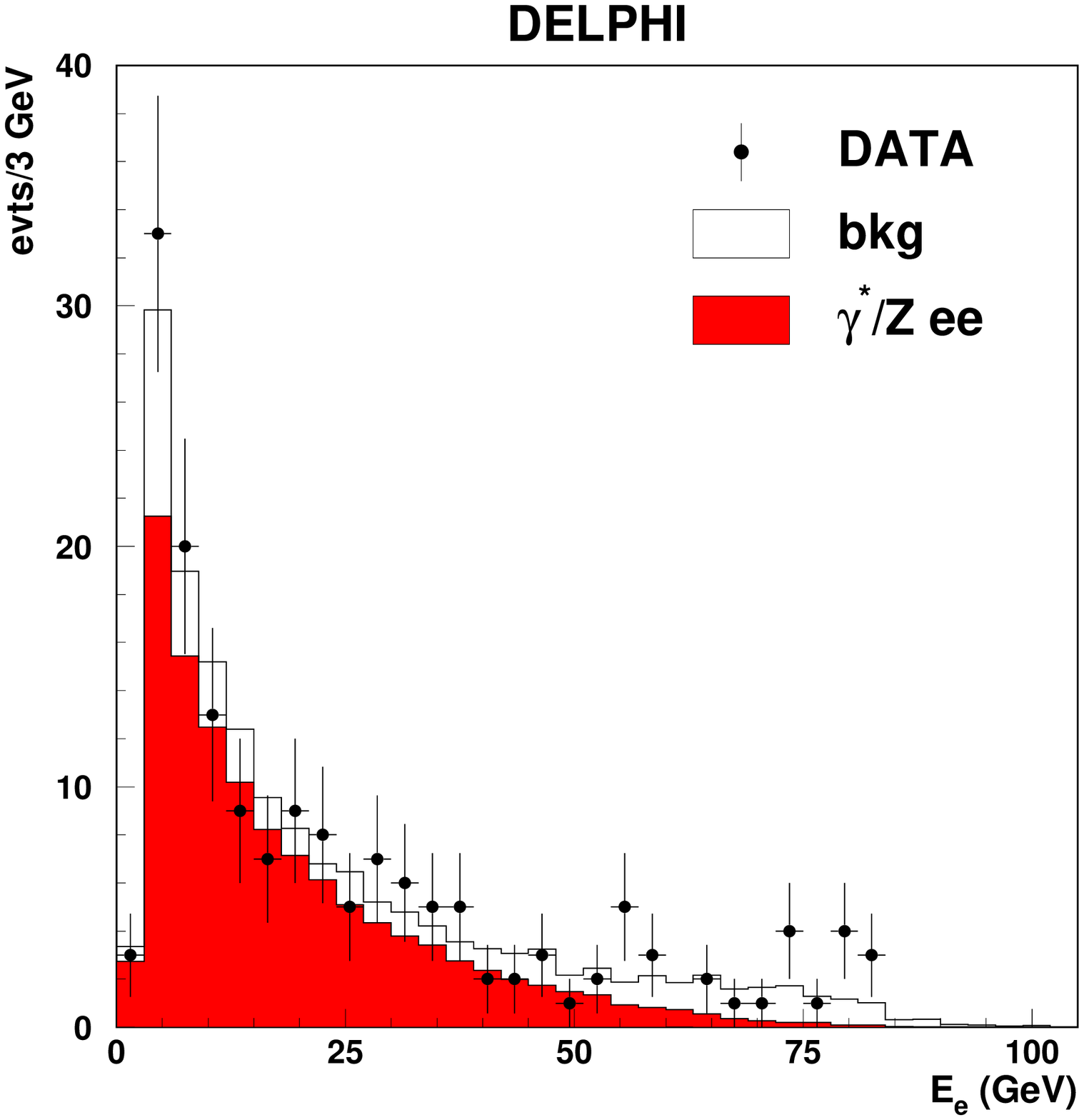}}  \\  
    \scalebox{0.5}{\epsfig{file=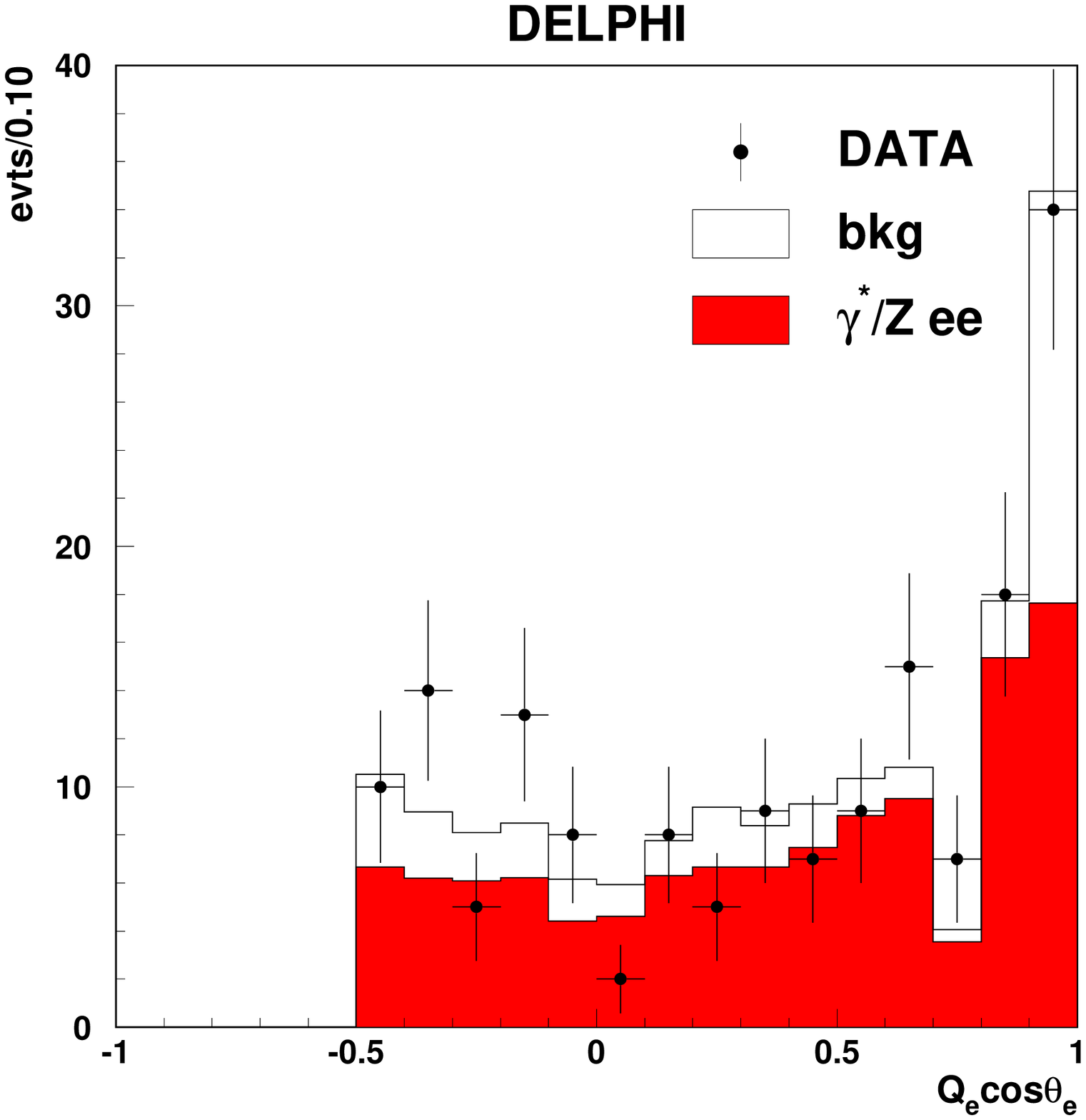}}
  \end{array}
  \)
\end{center}
\caption{$\gs/Zee$ channel summed over all centre-of-mass energies:
  energy spectrum {\it (top)} and signed angle $Q_{e} \cos
  \theta_{e}$, {\it (bottom)} of the tag electron 
  for hadronic and $\mm$ final states with  $m_{f\fbar} > 60$ GeV/$c^2$, 
  in the selected signal sample.
  The points with error bars
  represent real data, the histograms the simulation.}
\label{fig:ZeeEle}
\end{figure}

The distributions of the energy and of the signed angle, $Q_{e} \cos
\theta_{e}$, of the tag electron after the kinematic fit for hadronic
and $\mm$ events with  $m_{f\fbar} > 60$ GeV/$c^2$, are shown in
Figure~\ref{fig:ZeeEle} summed over all the centre-of-mass energies. 
The observed spectra are in agreement with the predictions from the
simulation.  
%
%
\subsection{Systematic uncertainties}

The measurement uncertainty is dominated by the limited real data statistics. 

In the hadronic channel three sources of systematic errors
were considered: the efficiency in the electron selection
procedure, the modelling of the contribution from 
two-photon events, which represents the largest background
component in the low invariant mass region, and the modelling of the
fragmentation in the $q\bar{q}(\gamma)$ contribution, which represents
the largest background component in the high invariant mass region.

The uncertainty on the efficiency of the electron identification was
estimated by comparing the number of selected events in the data and in
the simulation for a sample enriched in $WW$ events (about 85\% purity) 
with at least one of the two $W$'s decaying, directly or in cascade, into a final state
containing an electron. The same criteria for electron identification
and isolation were adopted as in the $Zee$ analysis, but the signal
selection criteria were changed to maximize the product efficiency
times purity of the $WW$ selection. Assuming the Standard Model
prediction for the $WW$ cross-section, including the ${\cal O} (\alpha)$
electroweak corrections via the so-called Leading Pole
Approximation~\cite{4fYR}, and attributing the entire 
discrepancy between the observed and the expected number of events to
the different electron identification efficiency in the data and in the simulation,
the relative difference in the efficiency was found to be $ \Delta
\varepsilon_{e}/\varepsilon_{e} = (-2.2\pm3.6)\%$ 
where the error accounts both for the data and the simulation statistics. 
The error on the difference was used for the computation of the
systematic error. 

The uncertainty in modelling of two-photon events could arise from 
the bad modelling either of the direct or of the resolved photon
contribution. As described in~\cite{4fmc}, in the region of single tag
the direct component was simulated using the WPHACT generator and
the resolved component using PYTHIA 6.143.
To match the direct and resolved components in the region $m_{q\bar{q}} < 40$
GeV/$c^2$ the WPHACT generator was run with constituent quark masses.
The direct component of single tag events with $m_{q\bar{q}} > 40$
GeV/$c^2$ was instead simulated with the WPHACT generator using
current algebra quark masses.  
To gauge the effect of the different quark masses for the single tag
low mass direct component, a fully simulated sample with current algebra quark
masses only was used to evaluate the effect both on signal efficiency
and on the background cross-section. The change in the quark mass does
not affect the $\gs/Zee$ signal at any stage of the selection, while
the background at the end of the selection is increased by about 5 fb
in each invariant mass region. 
Concerning the resolved photon component, 
at $\sqrt{s} = 200$ GeV the cross-section of this background amounts,
at the generator level, to about
10 fb in the $\gs ee$ signal region and 17 fb in the $Zee$ one.
A different generator, TWOGAM~\cite{twogam}, predicts background
cross-sections of 5 fb and 9 fb, respectively, in these two regions.
As the topologies of the resolved photon and $\gs/Zee$ events are similar, the
same selection efficiency was assumed for this background and the signal.
Therefore the difference between the PYTHIA and TWOGAM predictions at
the generator level, which is stable in the range $\sqrt{s} = 183-207$
GeV, was taken as systematic error on the measured cross-section.

The uncertainty in modelling the fragmentation and hadronization in
$q\bar{q} (\gamma)$ events was evaluated using a simulation sample
produced with the ARIADNE generator. The background cross-sections were
found to be larger, but within the statistical error, leading to a
decrease of the measured cross-sections of 3$\pm$5 fb in
the low invariant mass region and 7$\pm$13 fb in the high invariant mass one. The largest of the statistical errors from the ARIADNE and
PYTHIA samples, 3 fb in the $\gs ee$ signal region and 9 fb in the
$Zee$ one, were taken as systematic error. 

These three systematic uncertainties, together with the error on the luminosity, were
taken as fully correlated at the different centre-of-mass energies, while
the errors on the background cross-section and on the signal
efficiency due to the limited simulation statistics were considered
uncorrelated among the different energies. 

The contributions of the sources of systematic uncertainty in the
hadronic channel at 189 GeV are summarized in Table~\ref{tab:ZeeSyst}.
The total systematic uncertainty amounts to $\pm10\%$ in the region 
$15 < m_{q\qbar}<60$~GeV/$c^2$ and to $\pm6\%$ for  
$m_{q\qbar}>60$~GeV/$c^2$.

The contributions of the different sources of systematic errors in the
leptonic channel are summarized in Table~\ref{tab:ZeeSystmm}.
The main source of systematic error is the limited simulation
statistics, both for the signal and for the background.
The effect of the uncertainty on the efficiency of the electron 
identification was measured to be negligible
using relaxed identification criteria. 
The total systematic uncertainty amounts to about $\pm5\%$  
per energy point. 
Assuming no energy correlation of the systematic errors, 
the overall systematic uncertainty on the energy averaged cross-section 
was estimated to be  $\pm2.5\%$, an order 
of magnitude smaller than the statistical uncertainty.
\begin{table}[tb]
\begin{center}
\vspace{0.5cm}
\begin{tabular}{|l|c|c|}
\hline\hline
 Systematic effect  &   \multicolumn{2}{|c|}{Error on $\sigma$ (pb)}  \\
\hline \hline
& $15 < m_{q\qbar} < 60$~GeV/$c^2$ & $m_{q\qbar} > 60$
~GeV/$c^2$ \\
\hline
 $\Delta\varepsilon_e$                                 &  0.009 &  0.022 \\ 
\hline
 $\Delta\sigma_{bkg}$ ($\gamma \gamma$) direct         &  0.005  & 0.005 \\
 $\Delta\sigma_{bkg}$ ($\gamma \gamma$) resolved       &  0.005  & 0.008 \\
\hline
 $\Delta\sigma_{bkg}$ ($q\bar{q} \gamma$) fragmentation&  0.003  & 0.009 \\ 
\hline
 $\Delta\varepsilon$  due to simulation statistics         &  0.007  & 0.008 \\ 
 $\Delta\sigma_{bkg}$ due to simulation statistics         &  0.020  & 0.026 \\
\hline
 Luminosity  $\pm 0.6\%$                               &  0.002  & 0.005 \\
 \hline\hline
 Total                                                 &  0.024  & 0.038 \\
 \hline\hline
\end{tabular}
\end{center}
\caption{Contributions to the systematic uncertainty 
 on the $\gs/Zee$ cross-sections in the hadronic channel, in the two ranges
 of invariant mass of the hadronic system, at $\sqrt{s}=189$~GeV.
The selection efficiency, $\varepsilon$, is as
defined in table~\ref{tab:ZeePerf} 
and $\varepsilon_{e}$ is the electron identification efficiency.}
\label{tab:ZeeSyst}
\end {table}
\begin{table}[tb]
\begin{center}
\vspace{0.5cm}
\begin{tabular}{|l|c|}
\hline\hline
 Systematic effect  &  Error on $\sigma$ (fb)  \\
\hline \hline
 $\Delta\varepsilon$  due to simulation statistics      &  1.8  \\ 
 $\Delta\sigma_{bkg}$ due to simulation statistics     &  1.4  \\
\hline
 Luminosity  $\pm 0.6\%$                               &  0.7  \\
 \hline\hline
 Total                                                 &  2.4  \\
 \hline\hline
\end{tabular}
\end{center}
\caption{Contributions to the systematic uncertainty at $\sqrt{s}=189$~GeV
on the predicted $\gs/Zee$ cross-section in the leptonic channel. 
The selection efficiency, $\varepsilon$, is as
defined in table~\ref{tab:ZeePerf}.
The systematic errors were conservatively considered to be the same
for all centre-of-mass energies. The systematic due to the uncertainty
on the electron identification efficiency was measured to be negligible.}
\label{tab:ZeeSystmm}
\end {table}


\section{Combined single boson cross-sections}\label{sec:conclusions}

\begin{figure}[p]
\begin{center}
 \(
   \begin{array}{c}
    \scalebox{0.5}{\epsfig{file=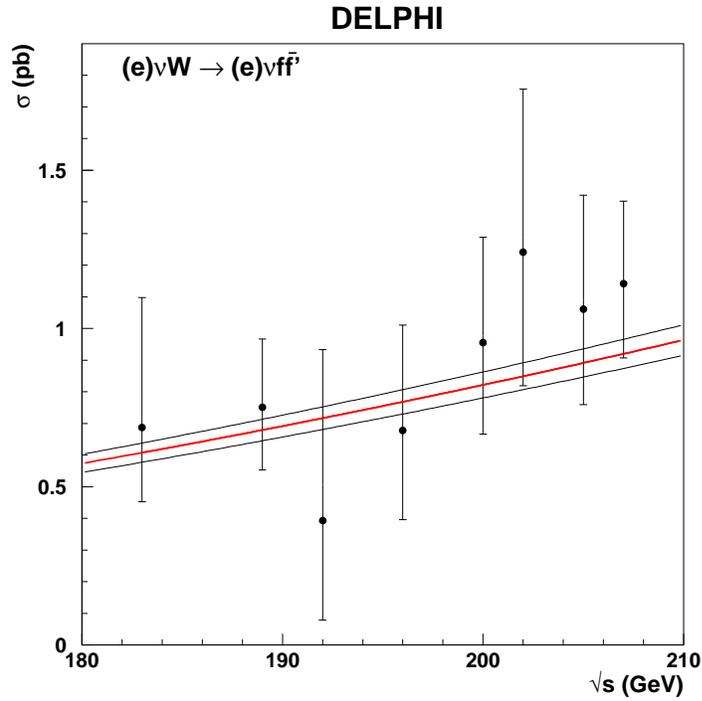}}\\
    \scalebox{0.5}{\epsfig{file=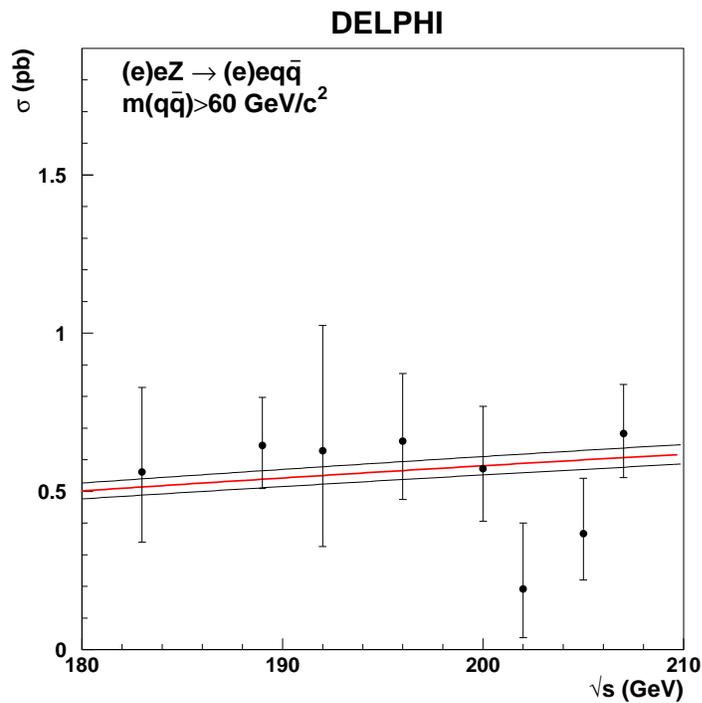}}
   \end{array}   
 \)
\end{center}
\caption{Cross-sections as a function of $\sqrt{s}$ for the
  $\wev$ channel {\it (top)} and  the $Zee$ channel
  $Zee \rightarrow e^+e^-q\bar{q}$ {\it (bottom)}.
  The solid curves are the Standard Model predictions, with a 5\% uncertainty
  band, computed with
  WPHACT~\cite{wphact}.}
\label{fig:xsec}
\end{figure}

The measured values for single boson cross-sections are compared with
the Standard Model predictions obtained with WPHACT~\cite{wphact} as a
function of the centre-of-mass energy. This dependency is shown in
Figure~\ref{fig:xsec} for single-$W$ and single-$Z$ production.
The theoretical uncertainty on the predictions amounts to 5\%.  
The overall compatibility with the Standard Model was checked by
considering the ratio $R$ of the measured to the predicted
cross-sections. At each energy point a Poissonian probability function
was constructed based on the number of observed events, the number of
expected background events and the signal extraction efficiency.  
A maximum likelihood fit to the data of the global probability function, being the
product over all probability functions for individual energies
convoluted with a multidimensional Gaussian describing the correlated
and uncorrelated systematic errors, was performed. The results were:

\begin{displaymath}
\begin{array}{cr}
R(e\nu_e qq') 
= 1.20 \pm 0.18~\mbox{(stat.)} \pm  0.14~\mbox{(syst.)}, \\ \\
R(e\nu_e \mu\nu_{\mu}) 
= 1.06 ^{+0.27}_{-0.25}~\mbox{(stat.)} \pm  0.03~\mbox{(syst.)}, \\ \\
R(e\nu_e e\bar{\nu}_e) 
= 1.07 ^{+0.38}_{-0.35}~\mbox{(stat.)} \pm  0.09~\mbox{(syst.)}, \\ \\
 R(ee q\bar{q}) 
= 1.22 ^{+0.17}_{-0.16}~\mbox{(stat.)} \pm  0.06~\mbox{(syst.)} & 15 <
m_{q\qbar} < 60~{\mathrm GeV}/c^2, \\ \\
 R(eeq\bar{q})
= 1.00 ^{+0.12}_{-0.11}~\mbox{(stat.)} \pm  0.05~\mbox{(syst.)} & m_{q\qbar} >
60~{\mathrm GeV}/c^2, \\ \\
 R(ee\mu\bar{\mu})
= 1.59 ^{+0.51}_{-0.43}~\mbox{(stat.)} \pm  0.03~\mbox{(syst.)}  & m_{\mu^+\mu^-} > 60~{\mathrm GeV}/c^2,
\end{array}
\end{displaymath}

\noindent where the systematic error represents only the experimental contribution.
The values found show a good agreement with the Standard Model predictions.
%
\subsection*{Acknowledgements}
\vskip 3 mm
 We are greatly indebted to our technical 
collaborators, to the members of the CERN-SL Division for the excellent 
performance of the LEP collider, and to the funding agencies for their
support in building and operating the DELPHI detector.\\
We acknowledge in particular the support of \\
Austrian Federal Ministry of Education, Science and Culture,
GZ 616.364/2-III/2a/98, \\
FNRS--FWO, Flanders Institute to encourage scientific and technological 
research in the industry (IWT), Belgium,  \\
FINEP, CNPq, CAPES, FUJB and FAPERJ, Brazil, \\
Czech Ministry of Industry and Trade, GA CR 202/99/1362,\\
Commission of the European Communities (DG XII), \\
Direction des Sciences de la Mati$\grave{\mbox{\rm e}}$re, CEA, France, \\
Bundesministerium f$\ddot{\mbox{\rm u}}$r Bildung, Wissenschaft, Forschung 
und Technologie, Germany,\\
General Secretariat for Research and Technology, Greece, \\
National Science Foundation (NWO) and Foundation for Research on Matter (FOM),
The Netherlands, \\
Norwegian Research Council,  \\
State Committee for Scientific Research, Poland, SPUB-M/CERN/PO3/DZ296/2000,
SPUB-M/CERN/PO3/DZ297/2000, 2P03B 104 19 and 2P03B 69 23(2002-2004)\\
FCT - Funda\c{c}\~ao para a Ci\^encia e Tecnologia, Portugal, \\
Vedecka grantova agentura MS SR, Slovakia, Nr. 95/5195/134, \\
Ministry of Science and Technology of the Republic of Slovenia, \\
CICYT, Spain, AEN99-0950 and AEN99-0761,  \\
The Swedish Research Council,      \\
Particle Physics and Astronomy Research Council, UK, \\
Department of Energy, USA, DE-FG02-01ER41155, \\
EEC RTN contract HPRN-CT-00292-2002. \\

\newpage


\newpage



\end{document}